\documentclass[journal]{IEEEtran}

\usepackage{tabularx,booktabs}
\usepackage{subcaption}

\ifCLASSINFOpdf
   \usepackage[pdftex]{graphicx}
\else
\fi
\usepackage{amsmath}
\usepackage{caption}
\usepackage{booktabs,siunitx}

\usepackage{subcaption}
\usepackage[ruled,vlined]{algorithm2e}
\begin{document}
%
\title{A Winograd-based Integrated Photonics Accelerator for Convolutional Neural Networks}
%
%
%

\author{Armin~Mehrabian, ~\IEEEmembership{Member,~IEEE,}
        Mario~Miscuglio,~\IEEEmembership{Member,~OSA,}
        Yousra~Alkabani,~\IEEEmembership{Member,~IEEE,}
        Volker J. Sorger,~\IEEEmembership{Senior~Member,~IEEE}
        and~Tarek El-Ghazawi,~\IEEEmembership{Fellow,~IEEE}

}
%
%

\markboth{Journal of Selected Topics in Quantum Electronics}%
{Shell \MakeLowercase{\textit{et al.}}: Bare Demo of IEEEtran.cls for IEEE Journals}
%

\IEEEpubid{\copyright~2019 IEEE}


\maketitle

\begin{abstract}

Neural Networks (NNs) have become the mainstream technology in the artificial intelligence (AI) renaissance over the past decade. Among different types of neural networks, convolutional neural networks (CNNs) have been widely adopted as they have achieved leading results in many fields such as computer vision and speech recognition. This success in part is due to the widespread availability of capable underlying hardware platforms. In parallel, hardware specialization can expose us to novel architectural solutions, which can outperform general purpose computers for the tasks at hand. Although different applications demand for different performance measures, they all share speed and energy efficiency as high priorities. Meanwhile, photonics processing has seen a resurgence due to its inherited high speed and low power nature. Here, we investigate the potential of using photonics in CNNs by proposing a CNN accelerator design based on Winograd filtering algorithm. Our evaluation results show that while a photonic accelerator can compete with current state-of-the-art electronic platforms in terms of both speed and power, it has the potential to improve the energy efficiency by up to three orders of magnitude.
\end{abstract}


%
\IEEEpeerreviewmaketitle

\section{Introduction}

%
%
%
%
\IEEEPARstart{T}{he} field of AI has undergone revolutionary progress over the past decade. Wide availability of data and cheaper than ever compute resources have contributed immensely to this growth. At the same time, advancements is the field of modern neural networks, known as deep learning (DL) have attracted the attention of academia and industry. This popularity is mainly owed to neural networks' success in a large gamut of AI applications including but not limited to computer vision, speech recognition, and natural language processing. Among the different types of neural networks, CNNs are considered the most viable architecture for AI applications. CNNs are remarkably versatile in most AI tasks. However, all of this comes at the price of high computational costs.\\

In the meantime, the use of integrated photonics in neural networks for implementing neuron functionalities has shaped to be an attainable alternative near future technology for limiting the power consumption and increasing the operating speed \cite{prucnal_neuromorphic_2017}\cite{chakraborty2018toward}\cite{feldmann2019all}. Photonics benefit from the coherent nature of electromagnetic waves, which interfere while propagate through a photonic integrated circuit (PIC). Central to many AI techniques and algorithms is implementation of hardware solutions that mimic the multiply and accumulate (MAC) function. The main advantage of photonic neural networks over electronics is that the energy consumption for performing a series of multiplications and additions does not scale with MAC speed. The training of an optical neural network necessitates an active modulation of the optical signal in a hybrid optical–electronic configuration \cite{george_neuromorphic_2019}. For this reason, these architectures face significant hurdles when compared to their electronic counterparts. To be competitive, they are expected to have low power consumption and high-speed electro-optic modulation \cite{wang_integrated_2018}\cite{liu_high-speed_2004}\cite{amin_0.52_2018}. Additionally, they require to pair with electrical to optical (EO), optical to electrical (OE) converters, and I/O interfaces. However, when trained, photonic neural networks do not rely on any additional energy for active switching. Therefore the architectures that perform tasks such as weighting, can be realized completely passive, and the computations happen without the consumption of any dynamic power \cite{bagherian2018chip}\cite{mehrabian2018pcnna}\cite{liu2019holylight}. In this panorama, all-optical neural networks (AONNs) represent a promising future. Current all-optical implementations in free space \cite{noauthor_optalysys_nodate} and in integrated photonics \cite{shen_deep_2017}\cite{hughes_training_2018}\cite{miscuglio_all-optical_2018} can outperform their electronic counterparts providing promises of great energy efficiency and speed enhancement for learning tasks.\\ \IEEEpubidadjcol

 In this manuscript, we explore the potentials of using high-speed, low-power photonics in a CNN accelerator by exploiting coherent all-optical matrix multiplication in wavelength division multiplexing (WDM), using microring resonator weight banks (MRRs). Our architecture is inspired by \cite{lavin2016fast}\cite{lu2017evaluating}, where Winograd filtering algorithm is adopted to perform convolution to speedup the execution time and reduce the computational complexity. We investigate the performance of our proposed architecture in terms of speed and power. Since our proposed architecture is analog at the core, we also investigate the robustness of neural networks executed on our proposed design in terms of tolerance against noise.\\
 
 We summarize the main contributions of this work as,
\begin{itemize}
    \item a first proposed photonic CNN architecture based on the Winograd filtering algorithm
    \item an analytical framework to evaluate the speed performance of our proposed accelerator
    \item an in-house simulator based on a modified Google Tensorflow tool to simulate the performance of our proposed photonic accelerator with power and noise awareness
    \item a modified training process to enhance robustness to inevitable hardware noise sources during the inference stage
\end{itemize}

\section{Convolutional Neural Networks (CNNs)}
\begin{figure}[ht]
    \centering
    \includegraphics[width=\linewidth]{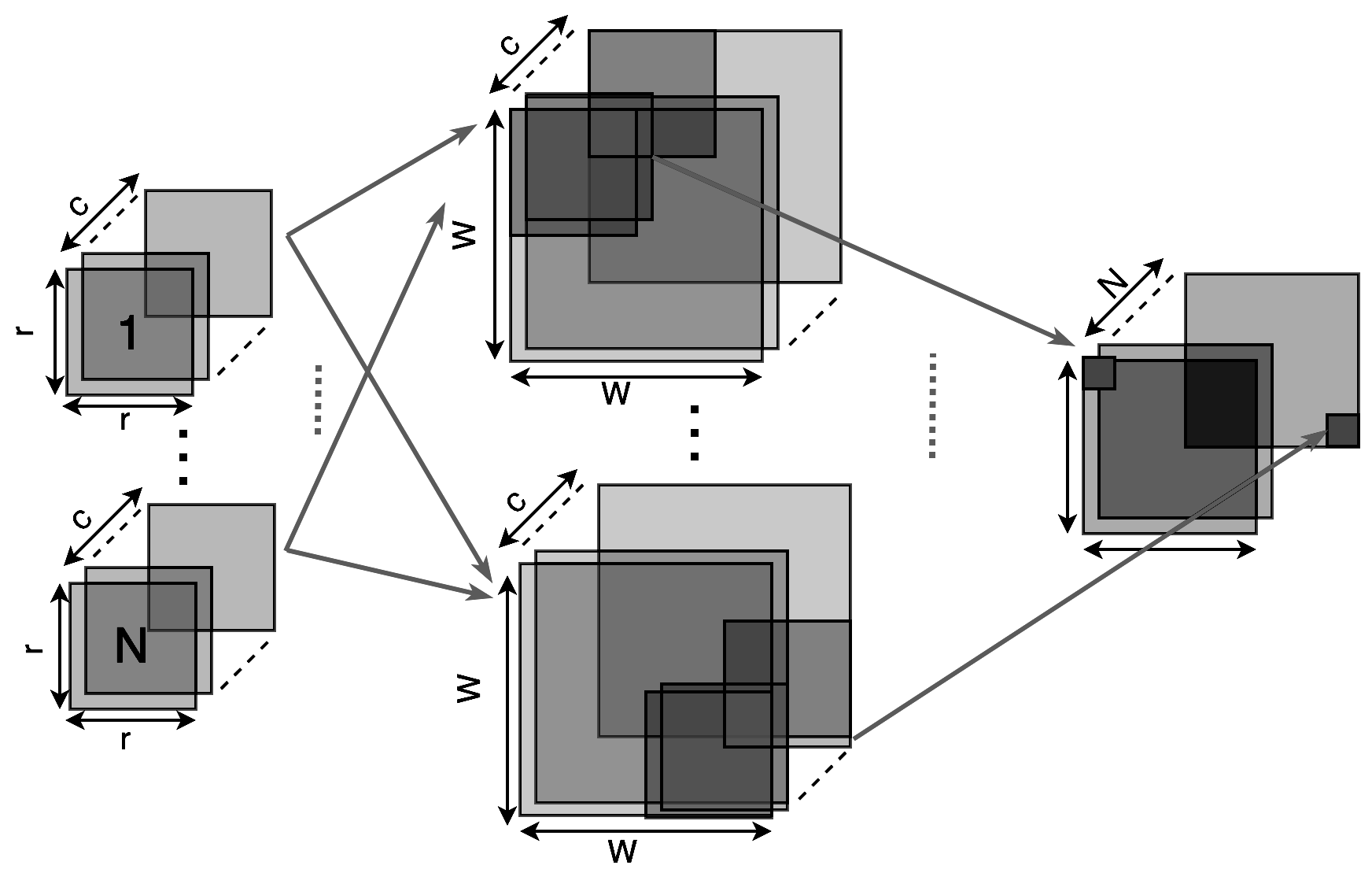}
    \caption{A single layer of a CNN. Each of the $N$ filters (left) scan the  input feature maps (middle) for features. This results in output feature maps, with $N$ channels equal to the total number of filters.} 
    \label{fig:CNN_sch}
\end{figure}

A CNN is a neural network comprised of one or more convolutional layers. CNNs are mostly known for their great performance on image data, however, their application extend to many other data types with local features. At the very high level, each convolution layer uses a collection of feature detectors, known as filters, that scan input data for presence or absence of a particular set of features. Hence, in a CNN layer, inputs and outputs are referred to as feature maps (fmap).  By cascading multiple of these convolutional layers, a hierarchy of feature detectors are formed. In this hierarchy, feature detectors closer to the input detect primitive features. As we move towards the final layers, the type of features detected become more abstract. Conventionally, the dimension of each filter in a a CNN is $3D$ with the two first dimensions being the height and width of the filer and the last dimension, known as the channel dimension, represents various filters. The use of  convolutional filters to scan input data had been practiced well before the rise of the field of deep learning and CNNs. However, in traditional signal processing, such filters are hand-engineered by experts, which can be costly, only designed for specific purposes, and vulnerable to designer bias. In a modern CNN, these filters are learned through the training process. Figure \ref{fig:CNN_sch} shows the overall architecture of a CNN layer.

\section{Photonic Realization of CNNs}\label{sec:photonic_realization}
In data communication and computation, photonics has the potential to offer practical solutions to overcome some of the limitations currently facing electronic systems. In a neuromorphic system, processing elements (PEs) are arranged in a distributed fashion with ideally large number of incoming (fan-in) and outgoing (fan-out) connections. Inspired by biological neural systems, some of these connection are required to connect neurons across farthest parts of the brain. In addition, neuromorphic PEs are in large part specific-purpose processors in contrast to the general purpose processors.\\

Neuromorphic processing can benefit from photonics in three major ways. First, photonics can significantly reduce the amount of energy consumed in interconnects among PEs by avoiding energy dissipation due to charging and discharging of electrical wires. Secondly, current neuromorphic algorithms known to neural networks, and in particular in CNNs, heavily rely on the multiply and accumulate (MAC) operation, which can be realized with very low energy budgets in photonics. Thirdly, photonics can increase communication and computation bandwidth by exploiting WDM. The adoption of WDM allow for higher density of computation and communication between PEs by packing more channels and parallel computations in a neuromorphic processor.

\subsection{Photonic Convolution Kernels and MAC Operation}
\begin{figure}[ht]
    \centering
    \includegraphics[scale=0.22]{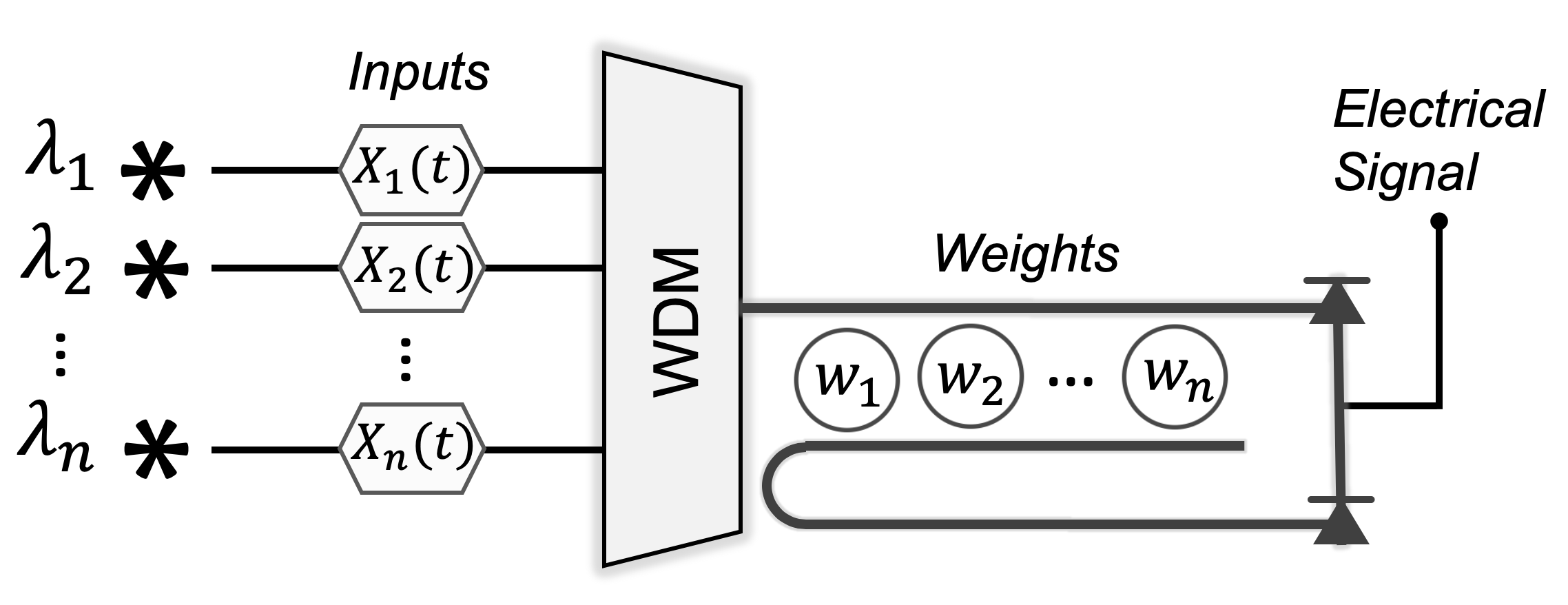}
    \caption{A broadcast-and-weight neuron. Inputs $X_i$  modulate different wavelength lasers. Modulated beams are then bundled through WDM.  } 
    \label{fig:BandW}
\end{figure}

One major advantage of a photonic MAC operation is that it can be performed with almost zero energy \cite{shen2017deep}. However, if the signal is converted from optical to electrical, the conversion and successive electronic manipulations impose additional energy loss. To build a photonic convolutional filter, we use a microring resonator (MRR) network proposed in \cite{tait2014broadcast}. Figure \ref{fig:BandW} depicts a single MRR neuron. In this schema input WDM signals are weighted through tunable MRRs. These weighted inputs are later incoherently summed up using a photodetector, which amounts to a MAC operation.\\

Thus, by the use of $N$ wavelengths, it is possible to establish up to $N^2$ independent connections. Maximum $N$ with current technologies is estimated to be around 108 channels resulting in a total of 10k connections \cite{tait2016microring}. It should be note that having closely spaced wavelengths as multiple laser sources while tuning the rings to match both resonance and FSR is a very challenging task. Although, on the source side, a set of phase-locked, equally spaced laser frequency lines can be generated using tunable optical resonators in a chip-based frequency comb generator \cite{hu2018single}. Moreover, on the MRR side, our system can leverage on Dense-WDM (DWDM). This is achievable due to strong optical confinement of silicon waveguides using tunable MRRs with more than $50 nm$ Free Spectral Range (FSR) with Quality factors Q close to $10^4$, which allow as many as $50$ channels \cite{xu2008silicon}.  Assuming approximately $0.8nm$ channel spacing, the resonance bandwidth can be broadened up to $0.4 nm$, while maintaining an estimated cross-talk level of $-10 dB$ \cite{chen2009integrated}.\\

Most of the modern neural networks have one or more fully-connected layers, which creates $N^2$ synaptic connections. On the other hand, $10k$ connections are barely sufficient to even implement miniature fully-connected neural networks and simple benchmark datasets such as MNIST with 728 neurons only in the input layer. In contrast, CNNs benefit from sparse connections between the local input regions and filters. A common CNN architecture usually has filters of shape $3\times 3$ up to $11\times 11$ that connect receptive fields and the filters. From functional point of view smaller filters are favored over larger filters, as they are capable of detecting finer local patterns. Larger and global patterns are detected in the layers closer to the output of CNNs. These features are more abstract and are built on top of the previously low-level features.\\

\begin{figure}[ht]
    \centering
    \includegraphics[width=.48\textwidth]{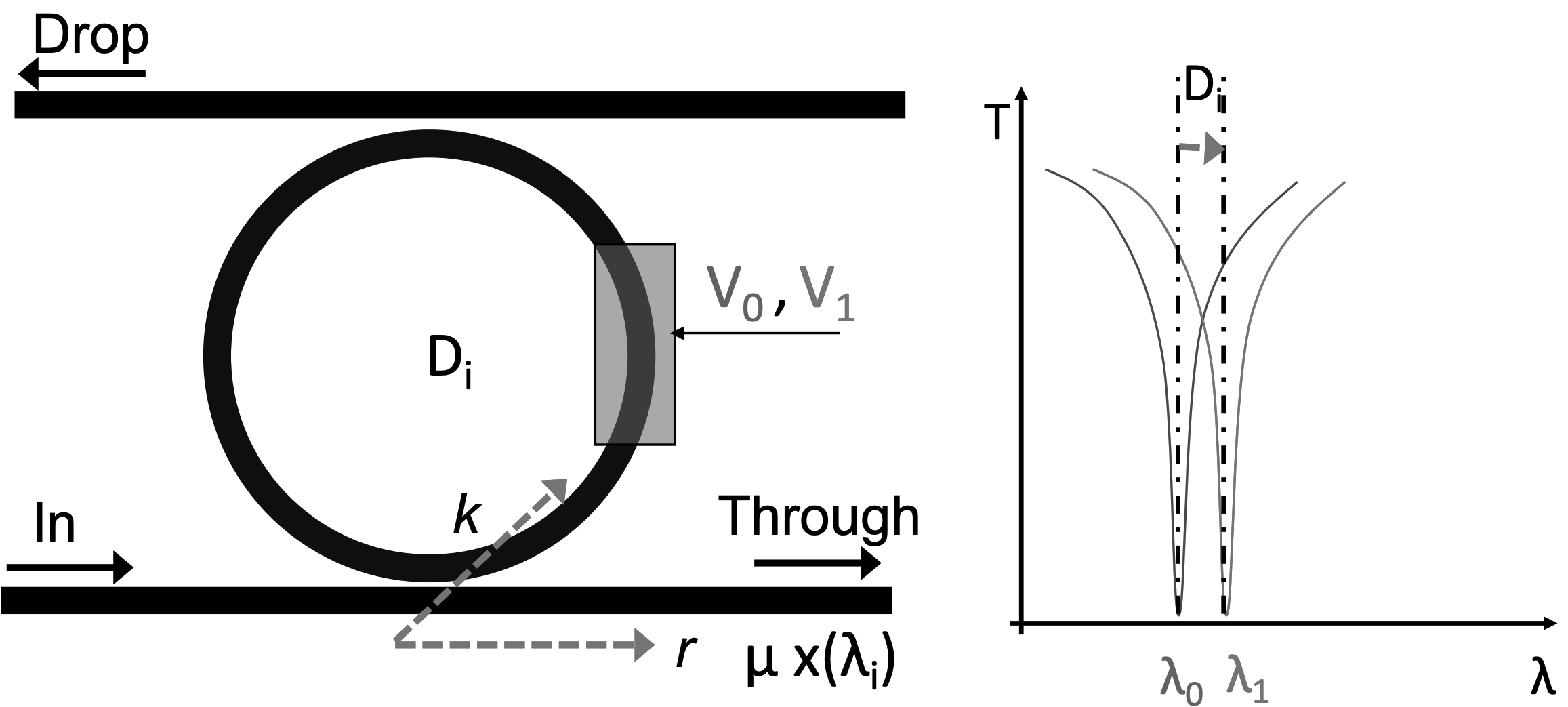}
    \caption{Microring resonator (MRR) operation for performing point-wise multiplication.} 
    \label{fig:MRR_1}
\end{figure}
We use the proposed scheme in Figure \ref{fig:BandW}, to perform two heuristic Winograd transformations and one element-wise matrix multiplication (EWMM) on each wavelength. Figure \ref{fig:MRR_1} shows the details of a MRR weighting function that operates on a single wavelength $\lambda_0$. The MRR acts as a tunable analog filter centred at $\lambda_0$, in which the voltage applied to the EOM module lets only a portion of light to travel through the waveguide. The modulation can be triggered by an analog electric field generated by a memristor. In this work we use a memristor device  which can store the weights with 6 bits of resolution \cite{stathopoulos2017multibit}. The transmission spectrum (T) of the ring has a natural resonant frequency of $\lambda_0$. When WDM light passes through the coupled waveguide, the component with wavelength $\lambda_0$ is coupled into the ring. By raising the bias voltage to V$_1$, the resonant frequency shifts to $\lambda_1$ due to the change in the effective refractive index of the ring. The difference between V$_0$ and V$_1$ controls the difference between $\lambda_0$ and $\lambda_1$, i.e. the transmission ($\Delta_i$). The variation of the transmission at $\lambda_0$ represents, in our scheme, the point-wise multiplication.\\

The most used MRR modulator has silicon based p-i-n junction that is side coupled to a wave guide as described in \cite{xu201723} or p-n junction reported \cite{ziebell_ten_2011}. Current silicon-based MRR modulators \cite{gardes_high-speed_2009}\cite{noauthor_osa_nodate}\cite{baba_50-gb/s_2013}, as well as foundry level implementations, exhibit a speed up to 50 GHz, with a driving voltage of usually a few Volts $(1-2V)$ and an efficiency (V$\pi$l) of few tenths of V·cm. Experimental results that corroborate our estimation are reported in \cite{dong_low_2009}, where Silicon-based electro-optic MRRs exhibit a modulation in a working spectrum of $0.1 nm$ and a speed of $11 GHz$ while have an insertion loss of as low as only $2dB$. This by no means is a limiting factor in the inference stage, considering that the network has been trained and the weights are set. Therefore, the latency of the network is given by the time-of-flight of the photon.\\

Beside the uncertainty due to fabrication imperfections, which could be compensated, the main source of noise that affects a MRR modulator is  electrical noise and, in this case, eventual non-ideality in setting the analog voltages with memristive device that could vary over time.  Moreover, for high data rate situations($>20Gb/s$), the intra-channel cross-talk becomes relevant, and power penalties need to be considered \cite{jayatilleka_crosstalk_2016}\cite{bahadori_crosstalk_2016}.
Regarding the operating dynamic power, the maximum allowed optical power flowing in each physical channel of the photonic accelerator is bound by the optical power that would produce nonlinearities in the silicon waveguides and the minimum power that the photo-detector can distinguish from noise (SNR=1). This sets the upper and lower operating range of photodiode, which we refer to as the dynamic range of the photodiode.
Foundry level \cite{noauthor_imec-epixfab_nodate} integrated Germanium photo-diodes can reach up to 40 GHz with a responsivity of $0.6 A/W$ and a Noise equivalent power (NEP) of around 1pW/$\sqrt{Hz}$ operating in reverse bias ($-2$V).
Research-level photodetectors working in the 100s of GHz range have also been demonstrated \cite{vivien_zero-bias_2012}\cite{salamin_100_2018}\cite{ma_100_2018}.
However, the  dynamic range of the photodiode needs to be accurately set to avoid saturation and account for the bit resolution \cite{shen_deep_2017}. For this scheme, according to the bit resolution, the estimated dynamic range is $20dB$.\\

The speed of the optical part of the accelerator, without considering the I/O interface, according to  \cite{tait_neuromorphic_2017} is given by the total number of the MRR and their pitch. Photodetection and phase cross-talk are expected to be the main sources of error in the proposed scheme.\\

Another issue in using MRRs is attributed to variations in device fabrications, which can result in the spectral shift of the resonance frequency. The resonance frequency of MRRs can be tuned in multiple way. Due to high optical sensitivity of materials such as silicon to temperature, thermal tuning is the most widely adopted tuning technique for MRRs. This can be achieved by placing micro-heaters on top of each MRR \cite{bogaerts2012silicon}\cite{xue2016thermal}. 

\subsection{Memristors as Analogue Weight Storage}

Neuromorphic systems inspired by human brain rely upon two major principles, namely massively distributed processing  and proximity of local memory to these processing elements. These memory units demand some level of programability (plasticity), with their programming speed requirements being in the $MHz$ regime. At this time, almost all state-of-the-art neural networks, perform the training and the inference phases separately. This means that once the weights are trained and set, for the inference phase, one does not need to change the weights. In addition, weights in our proposed system are represented by an analog voltage bias of MRRs. A potential weight storage for our system should be analog and non-volatile with long retention time.\\

Having said that, memristive memory devices have attracted the attention of researchers due to their interesting characteristics including but not limited to non-volatility, long state retention time, and ultra-low power consumption \cite{yoshida2007high}\cite{borghetti2010memristive}\cite{stathopoulos2017multibit}. Over the past few years the bit-resolution of such memristive memory devices has risen monotonically \cite{baek2004highly}\cite{merced2016repeatable}\cite{prakash2015resistance}\cite{lee2012multi}. Recently, authors in \cite{stathopoulos2017multibit} proposed, fabricated, and evaluated an analog multibit memristive memory with bit-resolution of up to $6.5$ bits. Each memristive device takes up $20\mu \times 20\mu$ in area and can retain the resistance state for up to $8$ hours. In \textit{AlexNet} the 3rd convolutional layer has the largest number of convolutional filter weights equal to $884,736$. Assuming overhead circuitry increases the footprint to approximately $50\mu \times 50\mu$, the memristive memory required for the largest layer of \textit{AlexNet} can be realized in less than $0.25 cm^2$.

\section{Fast algorithms for convolution operation} As the name suggest the convolution operation account for the bulk of all operations in a CNN during both training and inference stages. However, each of the training and inference stages demand a different type of performance requirement. During the training, the emphasis is more on throughput rather than time. This is mainly due to the fact that the model under train needs to observe a large "ensemble" of data, the batch, as fast as possible. Therefore, time is amortized over many inputs. On the other hand, during the inference stage, applications are mostly latency sensitive. For instance, in a self-driving car application only a few input image scenes are needed to be processed per second, but that is required to be at a very low latency timescale. Having said that, a neuromorphic processor designed for inference is expected to satisfy stringent timing requirements.\\

An important parameter that is shown to have a significant impact on the latency of CNNs is the size of their filters. It is generally known  from a functional point of view that CNNs with smaller filters are preferred over CNNs with large filters \cite{simonyan2014very}\cite{he2016deep}\cite{szegedy2016rethinking}. Table \ref{tab:CNN_filter_breakdown} shows the breakdown of filter size for some of the state of the art CNN architectures. This is mainly due to the fact that small filters are better in finding local features without sacrificing the resolution. More abstract and more global features can be detected in higher layers of a CNN built on previous local layer features. As we discussed in section \ref{sec:photonic_realization}, a physical implementation of photonic MRRs favors small size filters due to limited number of available wavelength bands. This synergy between functional and photonic realization of CNNs is the primary motivation behind this work.\\

\begin{table}[!t]
\renewcommand{\arraystretch}{1.3}
\caption{Kernel size breakdown in state-of-the-art CNNs. It can be seen that filter of size $5\times5$ comprise only a minute fraction of total filters.}
\label{table_example}
\centering
\begin{tabular}{c|cccc}
\toprule
CNN & $1\times1$ (\%) & $3\times3$ (\%) & Small 1D filters & $5\times5$ (\%)  \\
\midrule
GoogLeNet & 64.9 & 17.5 & 1.7 & 15.9 \\
Inception V3 & 43.2 & 17.9 & 35.7 & 3.2 \\
Inception V4 & 40.9 & 16.1 & 43 & 0 \\
MobileNet & 93.3 & 6.7 & 0 & 0 \\
ResNet50 & 68.5 & 29.6 & 1.9 & 0 \\
VGG16 & 0 & 100 & 0 & 0 \\
\bottomrule
\end{tabular}
\label{tab:CNN_filter_breakdown}

\end{table}
\begin{figure*}[ht]
    \centering
    \includegraphics[width=\linewidth]{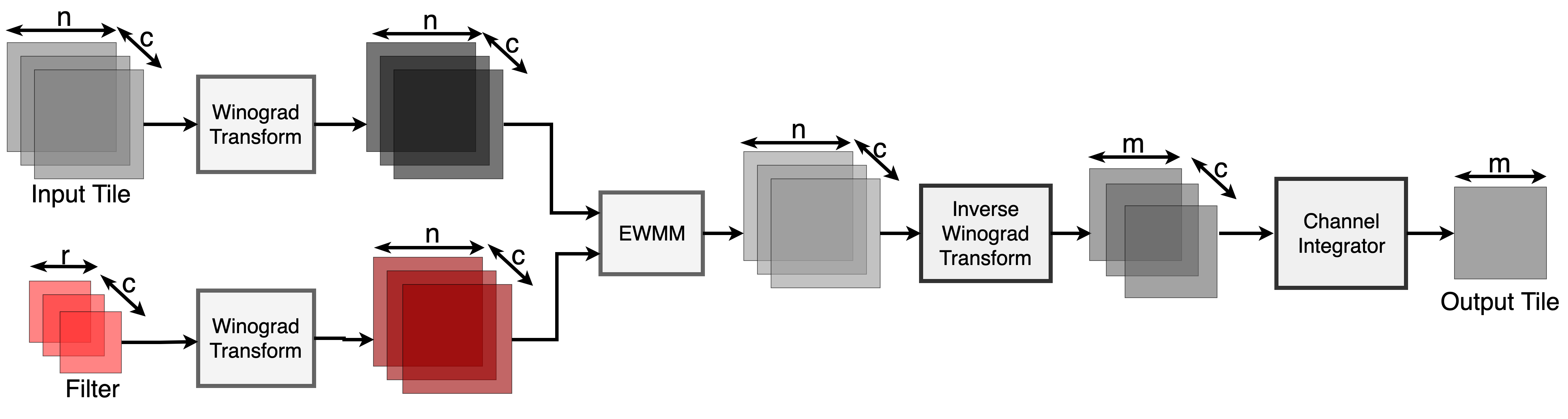}
    \caption{High-level flow diagram of Winograd filtering technique for convolution operation. Unlike conventional convolution, which computes a single output at a time, Winograd algorithm computes a tile of output, here of size $m\times m$. In order to generate an output tile, Winograd requires to fetch an input tile of size $n\times n$. Both input tile and filter are transformed into the Winograd domain. Within the Winograd domain, previously transformed input and filter are multiplied in an element-wise fashion. Finally, the output of the element-wise multiplication is transformed back into the original domain and channels are collapsed into a single value per tile element.} 
    \label{fig:winograd_pe}
\end{figure*}

At the time of writing this paper, there are three major ways to speed up the convolution operation. First, the General Matrix Multiplication (GEMM) approach, in which the convolution is converted to matrix multiplication operation using Toeplitz matrix. The downside to this method is that Toeplitz conversion expands the input by a factor of $r\times r$ where $r$ is the size of the filter. Second method uses Fast Fourier Transform (FFT) to perform tiled convolution operations. From Fourier theorem we know that cyclic convolution can be performed by transforming the input and filters into Fourier domain. An element-wise multiplication (also known as Hadamard multiplication) results in an equivalent of convolution, but in Fourier domain. An inverse FFT operation transforms the calculated convolution back into the original domain. FFT-based convolution had been the method of choice for convolution operation \cite{mathieu2013fast}\cite{vasilache2014fast}\cite{chetlur2014cudnn} until the recent past. Lately, it is shown that FFT-based convolution is better suited for larger filter sizes \cite{lavin2016fast}. The third method uses the Winograd filtering algorithm, which we explain in detail in the following section.
\subsection{Winograd Algorithm}
In a 2D convolution, a single output component of the convolution is calculated by,
\begin{equation}
    y_{n,k,p,q} = \sum_{n=1}^{c}\sum_{x=1}^{r}\sum_{y=1}^{r}x_{n,c,p+x,q+y}\times w_{k,c,x,y}
    \label{eq:convolution}
\end{equation}
The operation in equation \ref{eq:convolution} is repeated for all outputs convolution components. In a brute-force convolution the total number of multiplications required to perform a full convolution is equal to
\begin{equation}
    (m\times r)^2
    \label{eq:regular_complexity}
\end{equation}
where $m$ is the size of the output feature map channel and $r$ is the size of the filter. At the time of writing this paper, Winograd convolution in the most efficient convolution algorithm being used for CNNs \cite{lavin2016fast}. Winograd convolution is based on the minimal filtering principles. The algorithm states that in order to calculate $m$ outputs with a finite impulse response (FIR) filter of size $r$, denoted by $F(m,r)$, the number of required multiplications is, 
\begin{equation}
    n=(m+r-1)
    \label{eq:wino1d}
\end{equation}
While equation \ref{eq:wino1d} is derived for the 1D convolution operation, one can nest it with itself to acquire a 2D convolution. Therefore, the number of multiplications needed for the same 2D convolution is given by,
\begin{equation}
    (m+r-1)^2
    \label{eq:winograd_complexity}
\end{equation}

From equation \ref{eq:regular_complexity} and \ref{eq:winograd_complexity} we can infer that Winograd results in a reduction in the complexity by a factor of,

\begin{equation}
    \frac{(mr)^2}{(m+r-1)^2}
    \label{eq:complexity_reduction}
\end{equation}

 It should be noted that in our proposed photonic accelerator, multiplication operations are carried out by MRRs. Any reduction in the total number of multiplication operations, and thus MRRs, can save us not only in footprint of the design, but also in the design complexity.\\

Now, in order to understand how minimal Winograd works, let us first consider the case for 1D convolution. Let matrix $W$ be the matrix of weights, and matrix $D$ be the data matrix. Winograd computes the $F(2,1)$ convolution as following
\begin{equation}
    \begin{bmatrix}
    d_{0}       & d_{1} & d_{2} \\
    d_{1}       & d_{2} & d_{3} \\
\end{bmatrix}
\begin{bmatrix}
    w_{0}\\
    w_{1}\\
    w_{2}\\
\end{bmatrix}=
\begin{bmatrix}
    m_{1} + m_{2} + m_{3}\\
    m_{2}-m_{3}-m_{4}\\

\end{bmatrix}
\label{eq:1d_conv}
\end{equation}

where values $m_i$ are intermediate values found by
$$    m_{1} = (d_0-d_2)\times w_0 $$
$$    m_{2} = (d_1+d_2)\times \frac{w_0+w_1+w_2}{2} $$
$$    m_{3} = (d_2-d_1)\times \frac{w_0-w_1+w_2}{2} $$
$$    m_{4} = (d_0-d_3)\times w_2 $$

Above equations show that with only 4 multiplications between inputs and weights, Winograd can compute a $F(2,3)$ convolution. All $w_i$ terms can be pre-computed after the training stage. In order word, during the inference time, while data values $d_i$, corresponding to inputs change, the $w_i$ values remain the same throughout the inference. The 1D Winograd can be expressed by a closed matrix form as
\begin{equation}
    Y = A^T[(G\times w)\odot(B^T\times d)]
    \label{eq:1d_winograd_closed}
\end{equation}
where $A^T$, $B^T$, and $G$ are three heuristic transforms described by equations \ref{eq:A_T}, \ref{eq:B_T}, and \ref{eq:G}. $w$ is the weight vector and $d$ is the input vector.
\begin{equation}
A^T =\begin{bmatrix}
    1 & 1 & 1 & 0 \\
    0 & 1 & -1 & -1 \\
\end{bmatrix}
\label{eq:A_T}
\end{equation}

\begin{equation}
B^T =\begin{bmatrix}
    1 & 0 & -1 & 0 \\
    0 & 1 & 1 & 0 \\
    0 & -1 & 1 & 0 \\
    0 & 1 & 0 & -1 \\
\end{bmatrix}
\label{eq:B_T}
\end{equation}

\begin{equation}
G =\begin{bmatrix}
    1 & 0 & 0 \\
    \frac{1}{2} & \frac{1}{2} & \frac{1}{2} \\
    \frac{1}{2} & -\frac{1}{2} & \frac{1}{2} \\
    0  & 0 & 1 \\
\end{bmatrix}
\label{eq:G}
\end{equation}
One conclusion from equation \ref{eq:1d_conv} is that to compute a single output of 1D convolution only a window of $(m+r-1)$ input values are needed.\\

In a modern CNN the bulk of convolution operations are comprised of 2D convolutions. Equation \ref{eq:1d_winograd_closed} can be easily extrapolated for 2D convolution by nesting two 1D Winograd convolutions. The resulting 2D Winograd would be,
\begin{equation}
    Y = A^T[(G\times w \times G^T)\odot(B^T\times d)\times B]
    \label{eq:2d_winograd_closed}
\end{equation}
From \cite{lavin2016fast} for the case of $F(4\times 4, 3\times 3)$ matrices $B^T$, $G$, and $A^T$ have the forms,
\begin{equation}
A^T =\begin{bmatrix}
    1 & 1 & 1 & 1 & 1 & 0 \\
    0 & 1 & -1 & 2 & -2 & 0 \\
    0 & 1 & 1 & 4 & 4 & 0  \\
    0 & 1 & -1 & 8 & -8 & 0 \\
\end{bmatrix}
\end{equation}

\begin{equation}
B^T =\begin{bmatrix}
    4 & 0 & -5 & 0 & 1 & 0  \\
    0 & -4 & -4 & 1 & 1 & 0 \\
    0 & 4 & -4 & -1 & 1 & 0 \\
    0 & -2 & -1 & 2 &1 & 0 \\
    0 & 4 & 0 & -5 & 0 & 1 \\
\end{bmatrix}
\label{eq:B_T_2d}
\end{equation}

\begin{equation}
G =\begin{bmatrix}
    \frac{1}{4} & 0 & 0 \\
    \frac{-1}{6} & \frac{-1}{6} & \frac{-1}{6} \\
    \frac{-1}{6} & \frac{1}{6} & \frac{-1}{6} \\
    \frac{1}{24} & \frac{1}{12} & \frac{1}{6} \\
    \frac{1}{24} & \frac{-1}{12} & \frac{1}{6} \\
    0  & 0 & 1 \\
\end{bmatrix}
\label{eq:G_2D}
\end{equation}

The number of addition and multiplications required for Winograd transform, not the element-wise multiplications, increases quadratically with the tile size. Thus, Winograd is expected to perform most efficiently for smaller filter sizes, and thus smaller input tiles.

\begin{algorithm}
\SetAlgoLined
 \For{row=0; row$<$H; row+=m}{
  \For{column=0; column$<$H; column+=m}{
    \For{channel=0; channel$<$c;  channel+=1}{
        \For{filter=0; filter$<$N; kerne;+=1}{
        load input tile\;
        transform input tile\;
        load transformed filter\;
        perform EWMM
        }
    
    }
  }
  Output Winograd convolution result
 }
 \caption{Winograd for 2D convolution}
  \label{alg:winograd}
\end{algorithm}

\section{Architecture Design}
In this paper we propose a photonic CNN accelerator based on Winograd algorithm and realized using the photonic neuron introduced in \cite{tait2016microring}. Figure \ref{fig:winograd_pe} depicts the architecture of a single Winograd PE. Our proposed accelerator processes a single layer of a CNN at a time. This is mainly due to the fact that in a CNN different tiles of output feature maps are computed sequentially, and thus arrive at different times. But, in order to initiate processing of the next layer, all the inputs from the previous layer need to be available and synchronized. Our approach to process one layer at a time enforces this synchronization. Furthermore, implementing multiple layers of a CNN will result in large area overheads.\\

At the input of our accelerator, an input tile of shape $n \times n \times c$ along with filters of size $r\times r \times c$ are transformed into the Winograd domain. Input and filters' transforms are then multiplied element by element. The output of this multiplication needs to be transformed back using an inverse Winograd transform. The signals at this stage are digitized using an array of ADCs and placed onto the output line buffers to be stored back in the off-chip memory.\\

\begin{figure*}[ht]
    \centering
    \includegraphics[width=\linewidth]{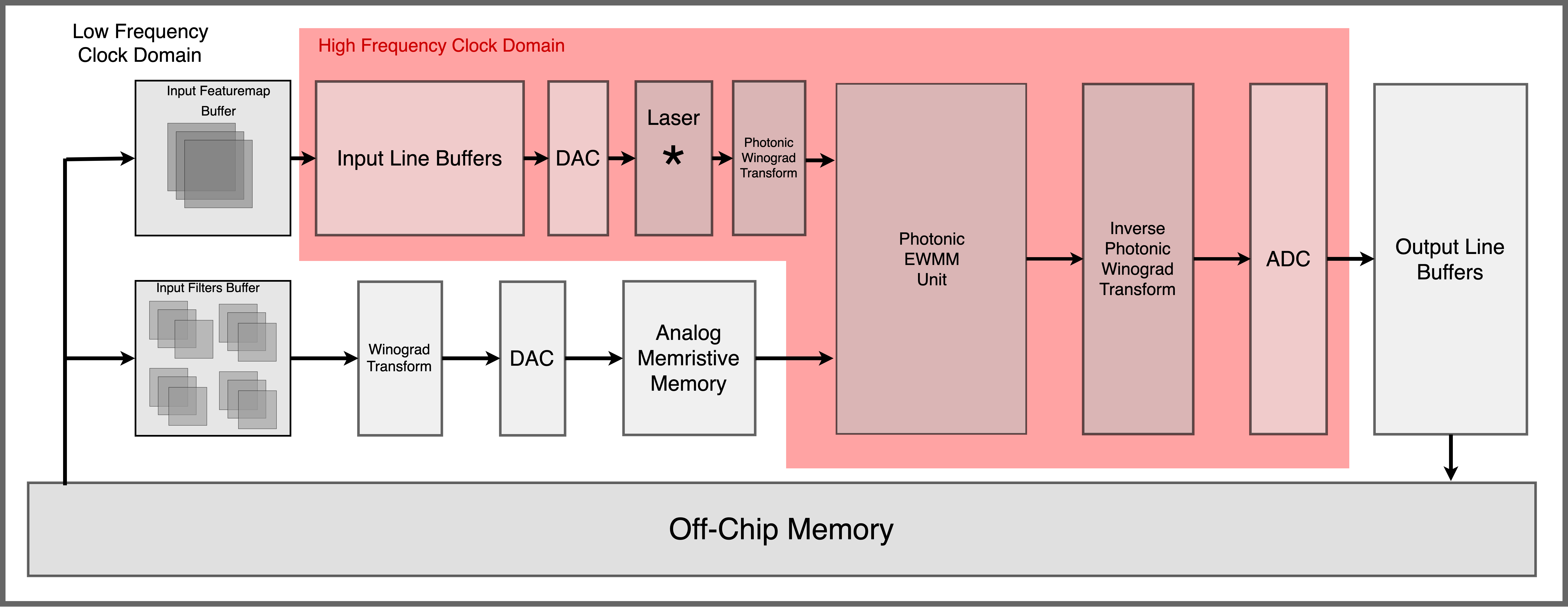}
    \caption{High-level architecture of our proposed photonic accelerator. Input feature-maps and filters are initially stored in an off-chip memory. Input tiles of size $n \times n$ are loaded into the input line buffer one at a time. Kernel weights do not change once the CNN is trained. Thus, we perform filters' Winograd in electronics and the cost is amortized over many input tiles. Winograd transform for input feature map tiles are computed in photonics. The photonic element-wise matrix multiplication (EWMM) unit performs the core Winograd element-wise Winograd multiplications. Outputs are digitized and placed onto the output line buffer. Finally, processed layer outputs are stored in the off-chip memory.} 
    \label{fig:winograd_arc}
\end{figure*}

Figure \ref{fig:winograd_arc} presents the overview of the our proposed architecture. Our proposed architecture runs on two clock domains. First a high speed $5GHz$ clock domain, which accommodates low latency components of the accelerator including the photonic components. In section \ref{sec:eval_speed} we explain our rationale on how we arrive at the $5GHz$ high speed clock frequency. The rest of the accelerator including input feature map buffers, filter buffers, filter Winograd DSP module, and filter path DAC run on a slower clock domain because there is no time sensitivity on filter path, and data transfer from/to off-chip memory. At the heart of our accelerator, we have an Element-Wise Matrix Multiplication unit, which we implement in photonics using photonic neurons. We store the input feature maps and filters in an off-chip memory. Both the input feature maps and filters require to go through Winograd transformation, which are matrix multiplications described in equations \ref{eq:B_T_2d} and \ref{eq:G_2D}. It should be noted that while input feature maps change for different tiles of inputs, filters are fixed for each layer. For that, we implemented the input feature map transformations in photonics and filter Winograd transformations in electronic DSP. This way, we will not pay the overhead associated with photonic implementations including the conversion of electronic filters to photonics. Later, the transformed filters and input feature map tiles are converted into analog signals to modulate the laser beams. However, as the filters are fixed over the processing time of the layer, analog filter signals need to be maintained for that time. Thus, we propose to use the non-volatile analog memristive memory bank, which maintains these voltages in their analog form for long retention times.\\

\begin{figure}[ht]
    \centering
    \includegraphics[width=\linewidth]{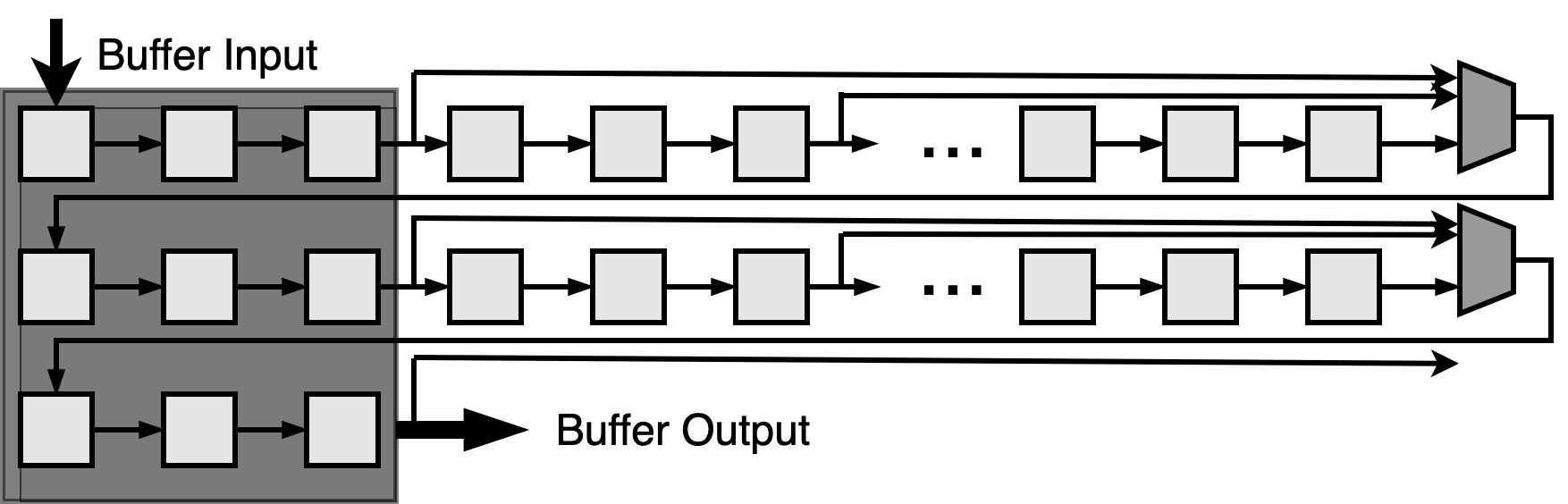}
    \caption{An example line buffer design to load and hold an input tile of size $3\times 3$.} 
    \label{fig:line_buffer}
\end{figure}

In Winograd convolution, and in each iteration, a tile of $n \times n$ is processed. In order to process an entire feature map, the transformed filter tile needs to move across the input feature map. In this paradigm two successive input tiles share size $(r-1)\times n$ elements. This, introduces data reuse opportunities, to avoid multiple queries of same data block. Here our goal is to exploit this opportunity at the front-end of our accelerator. Our design is inspired by the work in \cite{lu2017evaluating}, where authors utilize line buffers to avoid redundant queries from the off-chip memory. Figure \ref{fig:line_buffer} shows an example line buffer design to load and hold a $3\times 3$ input feature map tile. Input tiles are fetched from off-chip memory and loaded into the line buffer. Buffered tiles are then passed into the digital to analog converter (DAC) using parallel channels.\\

In parallel to the input data stream, transformed filter weights are converted into the analog signals to program the analog memristive memory. We then use the voltage generated using the stored analog signals to modulate the laser source for the filters. Each output signal generated by a DAC is then used to modulate a laser beam of a particular wavelength $\lambda_i$. It is worth noting that for each set of filters modulated by the laser source, input line goes through multiple iterations corresponding to different input tiles. Once both input tile laser beam and the filter laser beam are ready, the EWMM, multiplies each element of the Winograd input feature map tile with its corresponding Winograd filter value. The output from EWMM unit must be transformed back from Winograd domain into the original domain by the inverse Winograd transform. The result contains output feature map tiles for multiple channels $c$. Lastly, output feature map values are digitized and stored back into the off-chip memory.\\

A key principle in HPC is to try to minimize the IO and other communication latencies, compared to that of the computation time, to avoid unit under-utilization. From Algorithm \ref{alg:winograd} we can see that the two inner-most loops iterate over different channels of the input feature map tile and different filters. Moreover, operations within these two loops are independent form one another. This provides parallelization opportunities at the cost of additional hardware. In other words, the amount of parallelization and speed up we can achieve, scales linearly with the number of pipeline replications in our system. This linear scaling plateaus as soon as the computation bandwidth approaches the data transfer bandwidth. Our envisioned design uses an arbitrary number of 100 parallel paths. Our evaluation results in the next section justifies this selection.

\section{Evaluation}
In this section we evaluate the performance of our accelerator for the $3\times3$ filters of the \textit{VGG16} network against the recent FPGA \cite{zhang2015optimizing}\cite{lu2017evaluating}\cite{qiu2016going}\cite{suda2016throughput}\cite{zhang2018caffeine} and GPU implementations \cite{lu2017evaluating}.

\subsection{Speed}\label{sec:eval_speed}
Here we develop a model to estimate the execution time of the our accelerator. First we model the time required to convolve one input tile with one filter and we call it $T_{tile\_filter}$. Following that, we generalize the model to the case where we parallelize the process based on available resources.  For one input feature map tile and a filter, both, the input branch and the filter branch of the Figure \ref{fig:winograd_arc} are fully pipelined. Therefore, the execution time of a layer is determined by the longer of the two paths of filter path and input path. For each iteration of the filter path, input data path goes through multiple iterations. This is because a single filter operates on many input data tiles. That said, the input data path sets the upper bound on the delay. Our execution time model is comprised of two major components namely, the Input/Output time $(T_{IO})$ and the computation time $(T_{Comp})$. We define $(T_{IO})$ as.

\begin{equation}
    T_{IO}=max(T_{load},T_{offload})
    \label{eq:io_timing}
\end{equation}
where $T_{load}$ is the time it takes to transfer data from the off-chip memory to the input of the laser sources. Moreover, we can implement a total of $P_{input}$ DACs to speedup the data transfer. Considering the fact that our input matrices are of the shape $4\times 4$, we used an array of 16 DACs in this work. Similarly $T_{offload}$ is the time to store back the computed outputs from the inverse Winograd transform to the off-chip memory. The goal is to match the rate of the ADC at the output with input DAC to avoid any speed mismatch and thus congestion in the pipeline. At the time of this review both on-chip DACs and ADCs are capable of operating at sampling rates of more than 18 GS/s for bit resolution of at least 8 bits \cite{nazemi20153}\cite{xu201723}. Furthermore, with recent advances in memory technology, current memories are able to transfer data at high IO bandwidths up to more than $512Gb/s$ \cite{lee20151}. This high memory bandwidth allows us to buffer data and filters from off-chip memory at high transfer rates and feed it to our input line buffers. However, for our line buffers we need memories with high clock frequency and short access time. Current reported memory technologies have access time as short as $200ps$.\\

At the photonic core of our accelerator, $T_{compute}$ is,
\begin{equation}
    T_{compute}=T_{laser}+T_{Winograd}+T_{EWMM}+T_{iWinograd}
\end{equation}
where $T_{Winograd}$ is the time to compute the Winograd transform, $T_{EWMM}$ is the time to perform the element-wise matrix multiplication, and $T_{iWinograd}$ is the time compute the inverse Winograd transform. Once the laser is set up, input signals only incur a time delay equivalent to flight time of the light before they are fed into the ADC. Having said that the clock frequency of the pipeline is determined by
\begin{equation}
     frequency_{clock} \leq	 \frac{1}{min(T_{load}, T_{offload}, T_{compute})}
     \label{eq:clock_requirement}
\end{equation}
As a result of equation \ref{eq:clock_requirement}, we picked a clock frequency of $5Ghz$, which satisfies equation  \ref{eq:clock_requirement}. From equation \ref{eq:clock_requirement} $T_{tile\_filter}$ is simply found by,
\begin{equation}
    T_{tile\_filter} = \frac{1}{5GHz} = 200ps
    \label{eq:tile_time}
\end{equation}

\begin{figure}[ht]
    \centering
    \includegraphics[width=\linewidth]{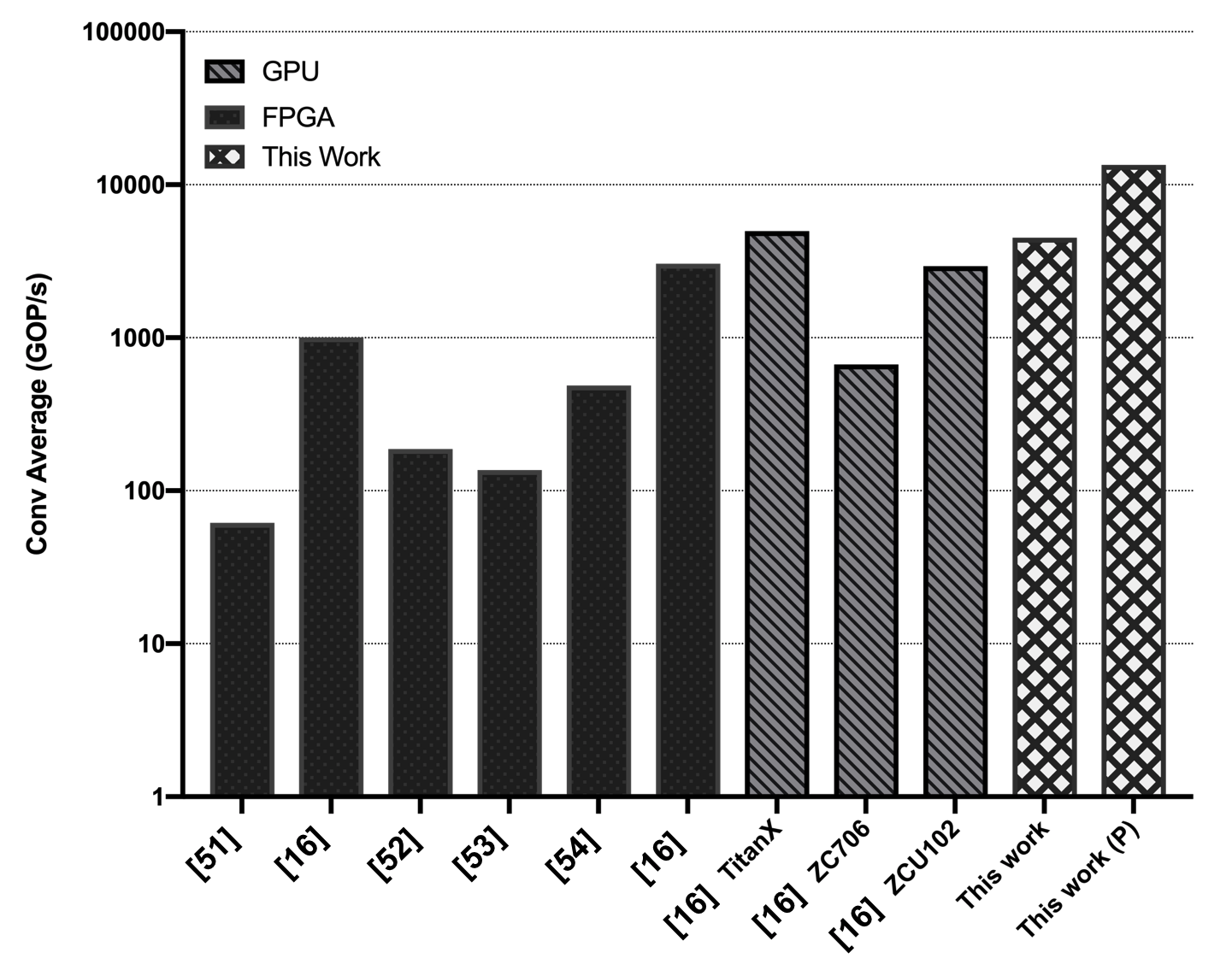}
    \caption{Comparison of convolution operation speed for FPGA, GPU, and our photonic implementation. The last column labeled with (p) represents the speed of the photonic core in the absence of electronics.} 
    \label{fig:speed_comp}
\end{figure}

For a $F(4\times 4, 3\times 3)$ Winograd, each $T_{tile\_filter}$ returns an output block of size 9 equivalent to 9 convolution operations. Having a clock frequency of $5GHz$, our proposed accelerator performs at $9\times 5G=45GOP/s$. Figure \ref{fig:speed_comp} shows the average convolution speed comparison of our proposed accelerator against the state-of-the-art FPGA and GPU implementations.



\subsection{Power}
\begin{figure}[ht]
    \centering
    \includegraphics[scale=0.27]{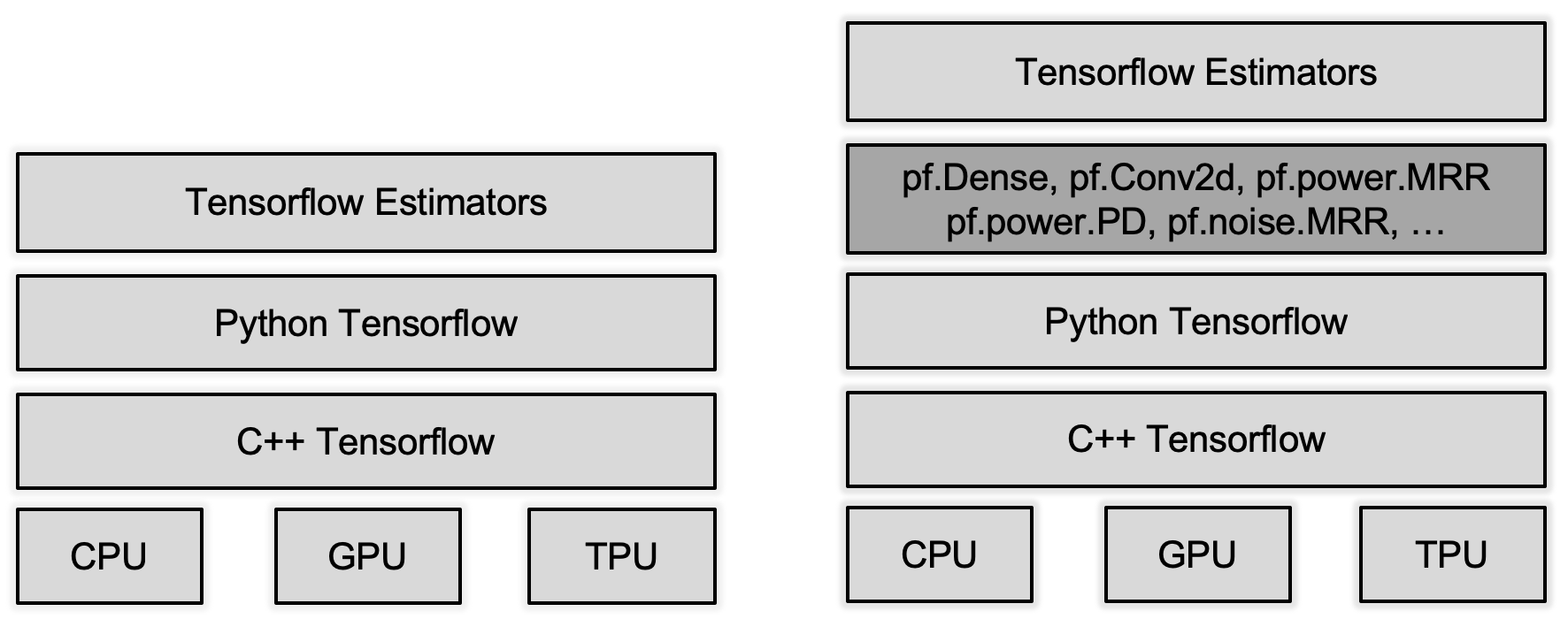}
    \caption{High level Tensorflow toolkit hierarchy vs. augmented Tensorflow.} 
    \label{fig:tensorflow}
\end{figure}

In order to estimate the dynamic power consumption of our proposed system, we built our in-house estimator by augmenting the standard Google Tensorflow tool. While primarily used for training and inference stages of neural networks, at the core, Tensorflow is a symbolic mathematical graph processing platform. Tensorflow enables users to express arbitrary computations into a dataflow graph, which is extremely useful in the context of neural networks. However, out-of-the-box Tensorflow is completely agnostic to physical realization of the neural networks being implemented. Thus, we augmented Tensorflow high-level API with mathematical models of electro-optical components. Figure \ref{fig:tensorflow} depicts the native Tensorflow toolkit hierarchy against our augmented version.
In our estimator, each primitive mathematical operation is given two physical models namely, the power model and the noise model. While, the noise model can impact the functionality, thus accuracy, in a neural network, the power model only models/measures consumed power. Table \ref{tab:tensorflow_mapping} shows some of these mathematical operations mapped to to their physical realizations.\\
\begin{table}[!t]
\renewcommand{\arraystretch}{1.3}
\caption{Mapping of primitive math operations to their hardware realization.}
\label{table_example}
\centering
\begin{tabular}{c|c}
\toprule
Math Operation & Photonic Representation  \\
\midrule
 Addition & Photodiode \\
 Multiplication & MRR \\
 Connection & Waveguide \\
 Non-linear Activation & Electro-absorption Modulator\\
\bottomrule
\end{tabular}
\label{tab:tensorflow_mapping}

\end{table}

Photodiode power can be simply derived from its Responsivity equation:
\begin{equation}
    R=\frac{I_{ph}}{P_{in}}=\lambda\frac{q}{hc}\eta \;\;\;\; [\dfrac{A}{W}]
\end{equation}
where $I_{ph}$ is the photocurrent, $P_{in}$ is the optical input signal power, $q$ is the electron charge, $\lambda$ is the wavelength, $h$ is the Planck's constant, and $c$ is the speed of light. It should be noted that the signal is encoded in the optical input power $P_{in}$. In this work use Aim photonics PDK values in \cite{aimspdk2019}. Similarly, we modeled both thermal noise and the shot noise in photodiode.
\begin{equation}
    I_{sn}=\sqrt{2q(I_{ph}+I_D)\Delta f}
\end{equation}
\begin{equation}
    I_{tn}=\sqrt{\frac{4K_BT\Delta f}{R_{SH}}}
\end{equation}
where $I_D$ is the dark current of the photodetector, $\Delta f$ is the noise measurement bandwidth, $K_B$ is the Boltzmann Constant, $T$ is temperature in Kelvins and $R_{SH}$ is total equivalent shunt resistance of the photodiode. For MRRs we accounted for per unit length propagation loss.

\begin{figure}[ht]
    \centering
    \includegraphics[width=\linewidth]{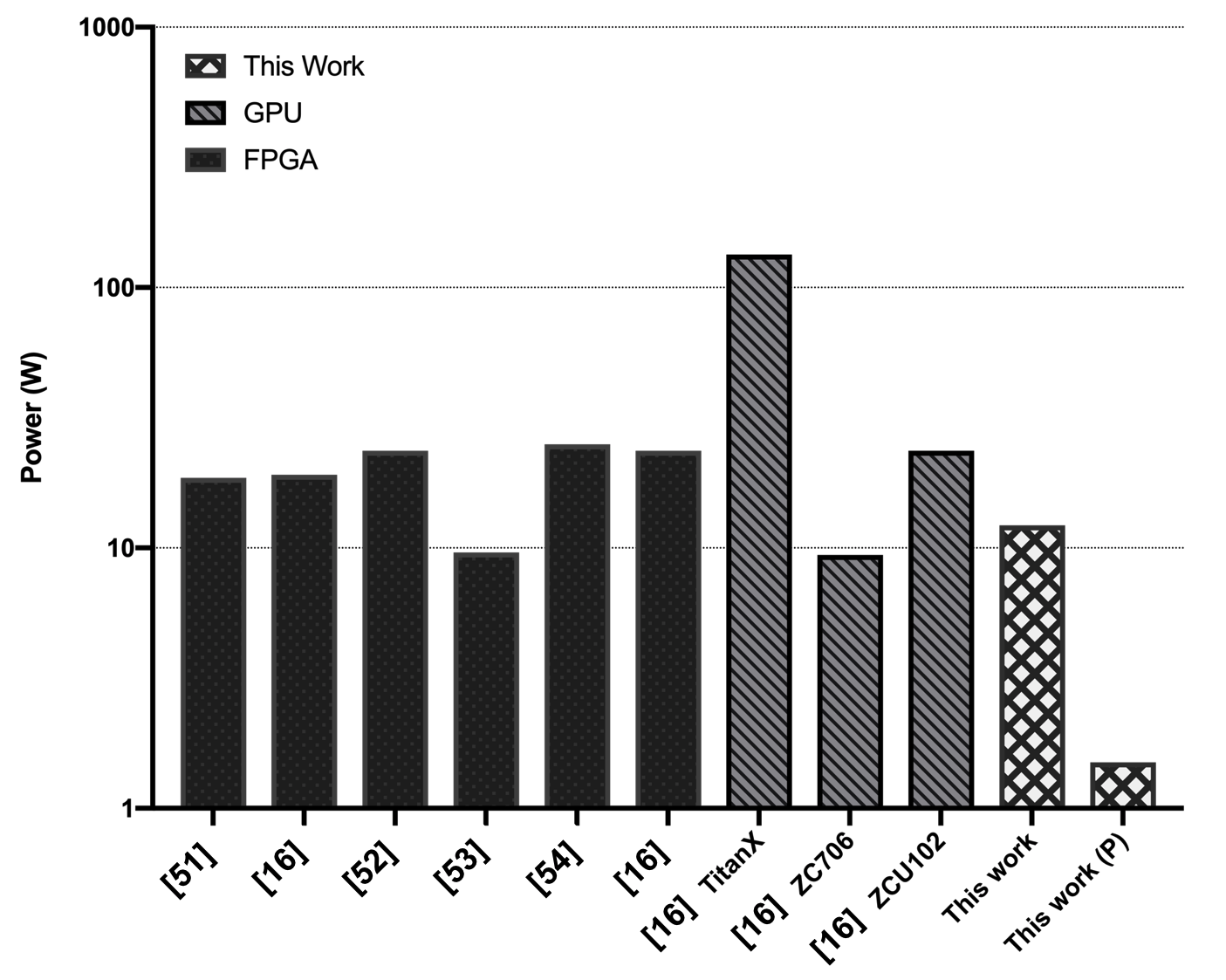}
    \caption{Comparison of convolution operation power for FPGA, GPU, and our photonic implementation. The last column labeled with (p) represents the power consumption of the photonic core in the absence of electronics.} 
    \label{fig:power_com}
\end{figure}
\begin{figure}[ht]
    \centering
    \includegraphics[width=\linewidth]{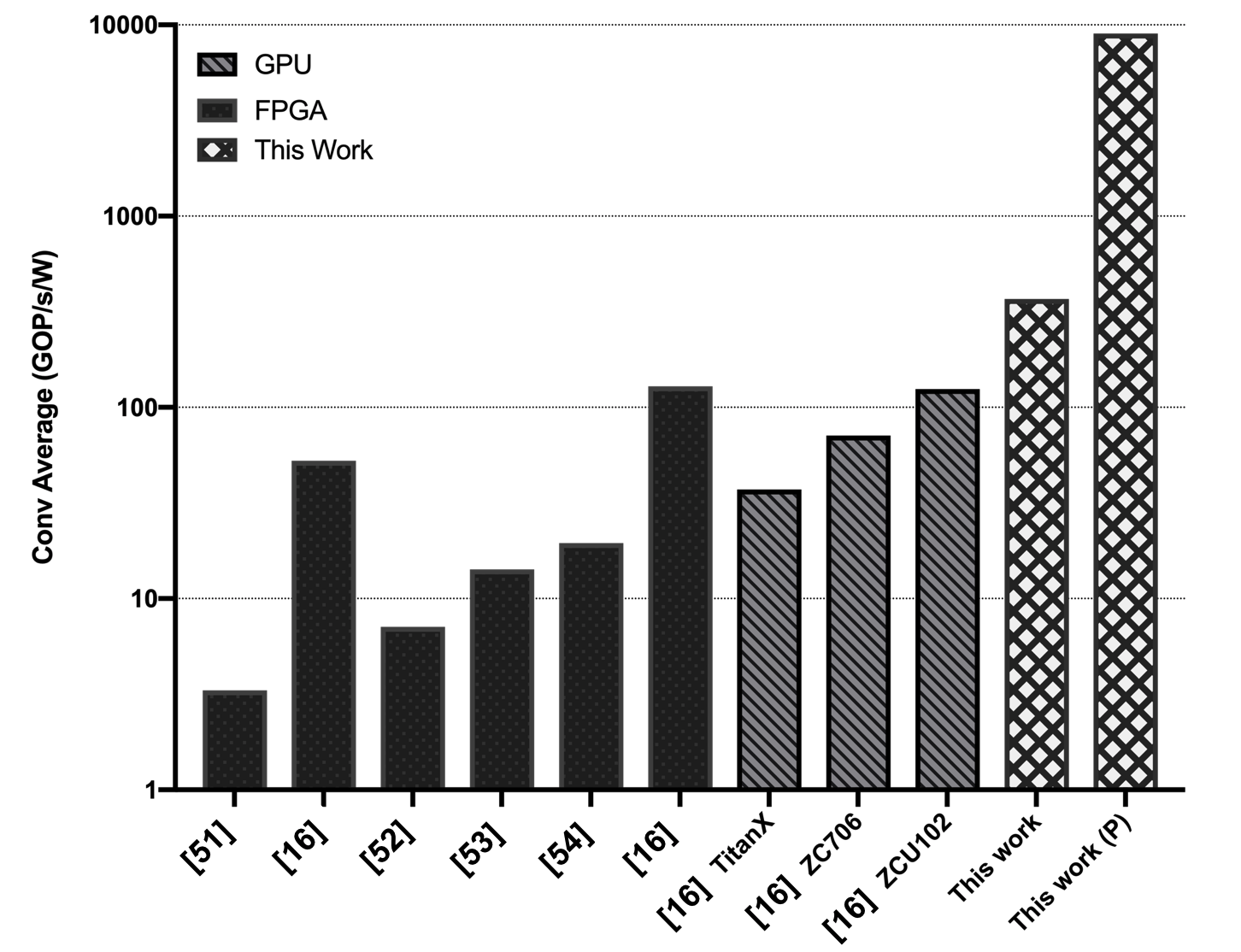}
    \caption{Comparison of energy efficiency for FPGA, GPU, and our photonic implementation. The last column labeled with (p) represents the power consumption of the photonic core in the absence of electronics. The results show that using photonics as an accelerator has the potential of improving energy efficiency by up to more than three orders of magnitude.} 
    \label{fig:efficiency}
\end{figure}

Figure \ref{fig:power_com} depicts the power comparison results. Finally We plotted the energy efficiency figure of merit defined by the ratio of speed to power in Figure \ref{fig:efficiency}.

\begin{figure}[ht]
    \centering
    \includegraphics[scale=0.4]{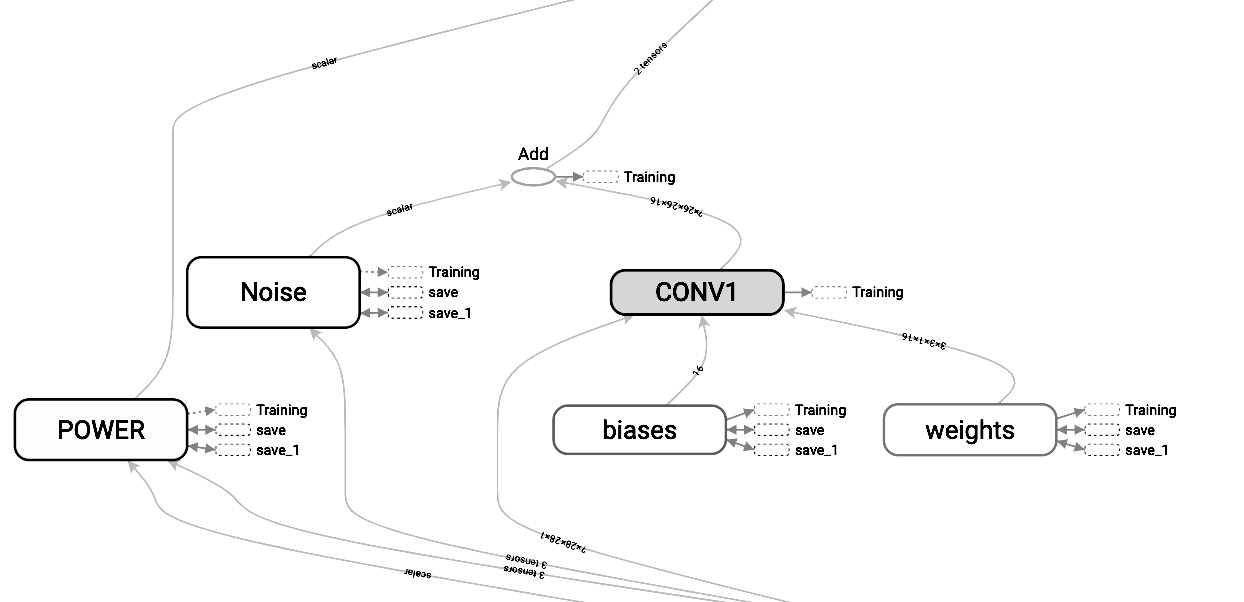}
    \caption{Visualalization of an augmented convolutional layer using power and noise models for \textit{VGG16} network.} 
    \label{fig:conv}
\end{figure}

\section{Training, Inference, and Noise} 
\begin{figure}
     \centering
     \begin{subfigure}[b]{0.5\textwidth}
         \centering
         \includegraphics[width=\textwidth]{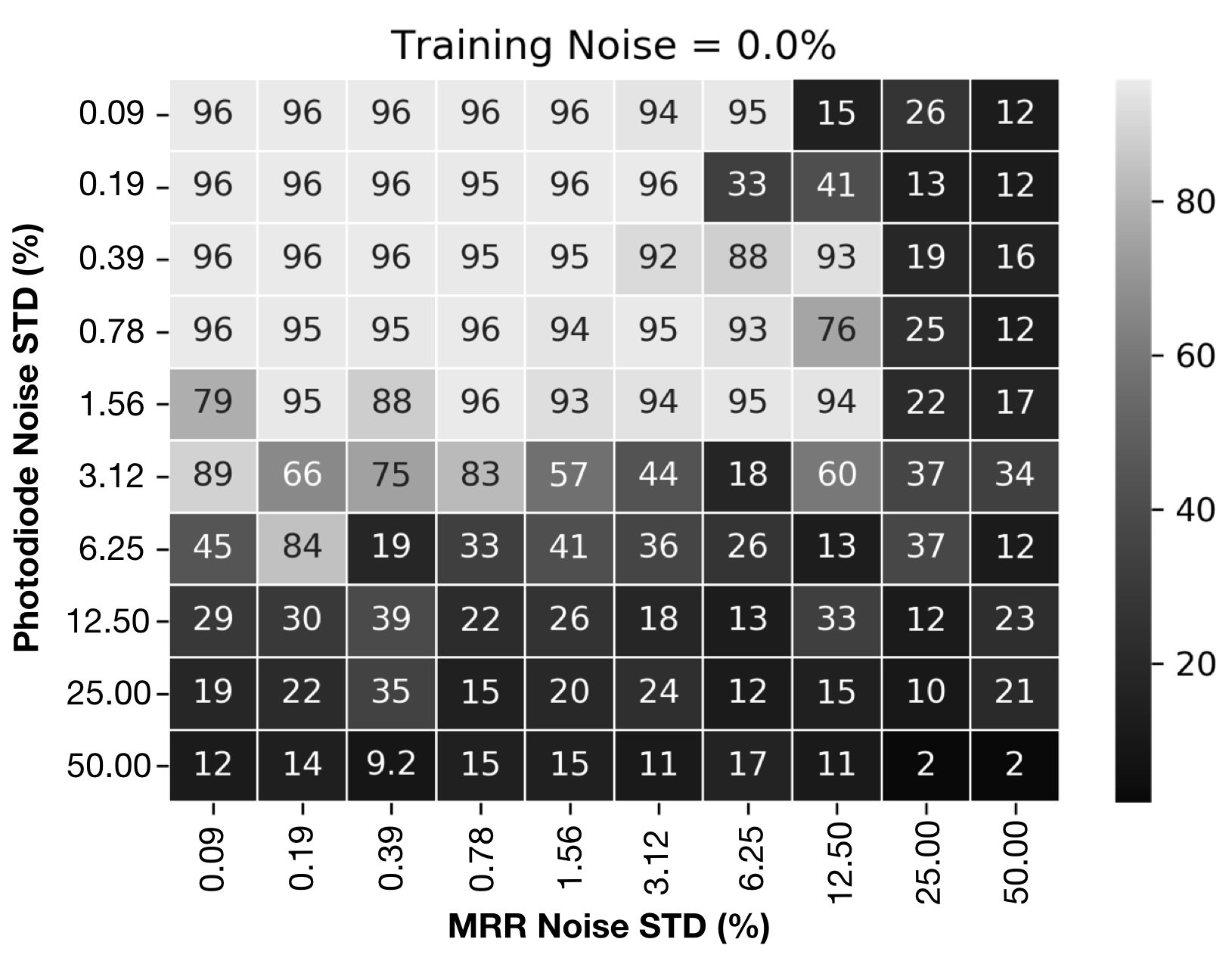}
         \label{fig:no_train_noise}
     \end{subfigure}
     \hfill
     \begin{subfigure}[b]{0.5\textwidth}
         \centering
         \includegraphics[width=\textwidth]{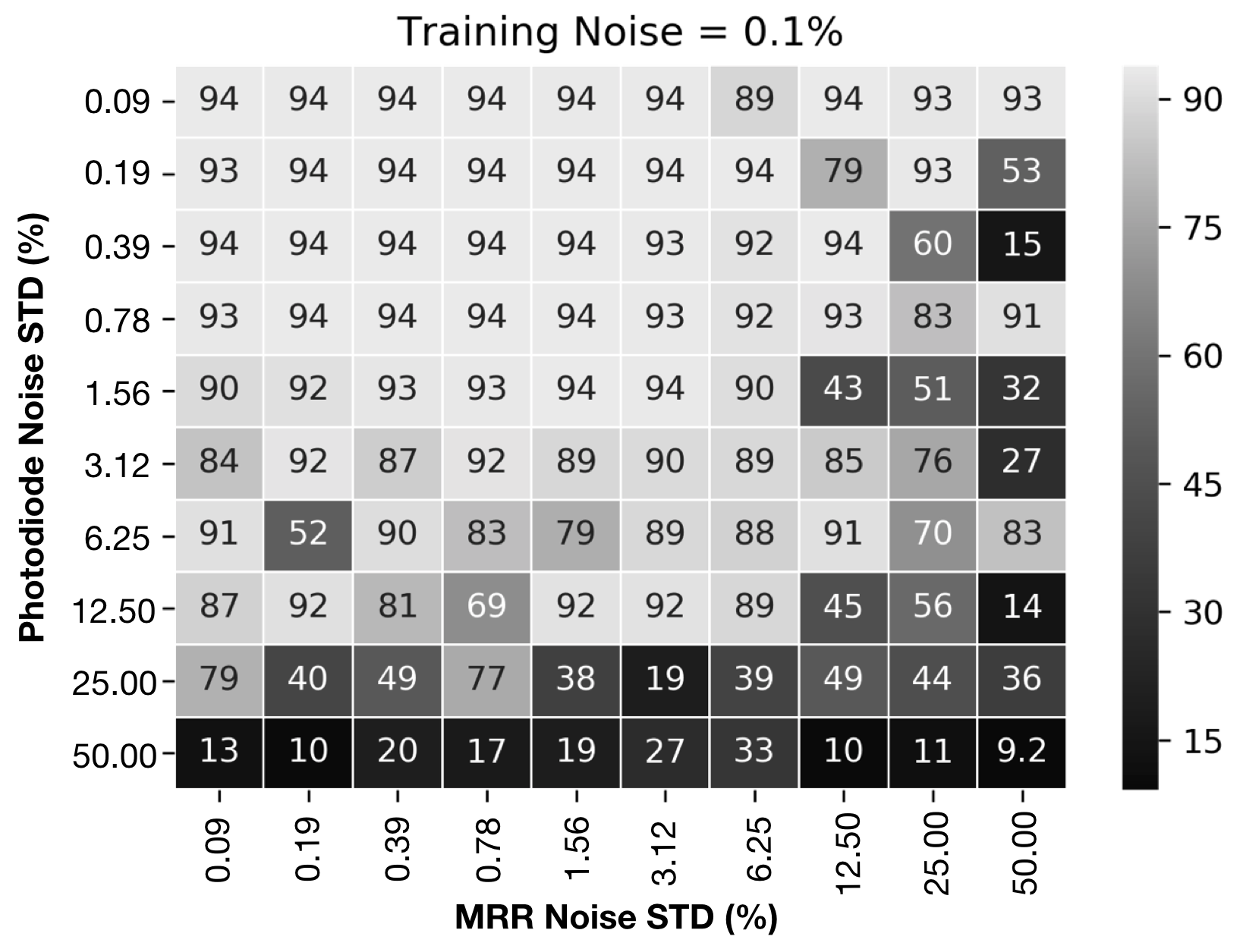}
         
         \label{fig:noise_001}
     \end{subfigure}
     \hfill
     \begin{subfigure}[b]{0.5\textwidth}
         \centering
         \includegraphics[width=\textwidth]{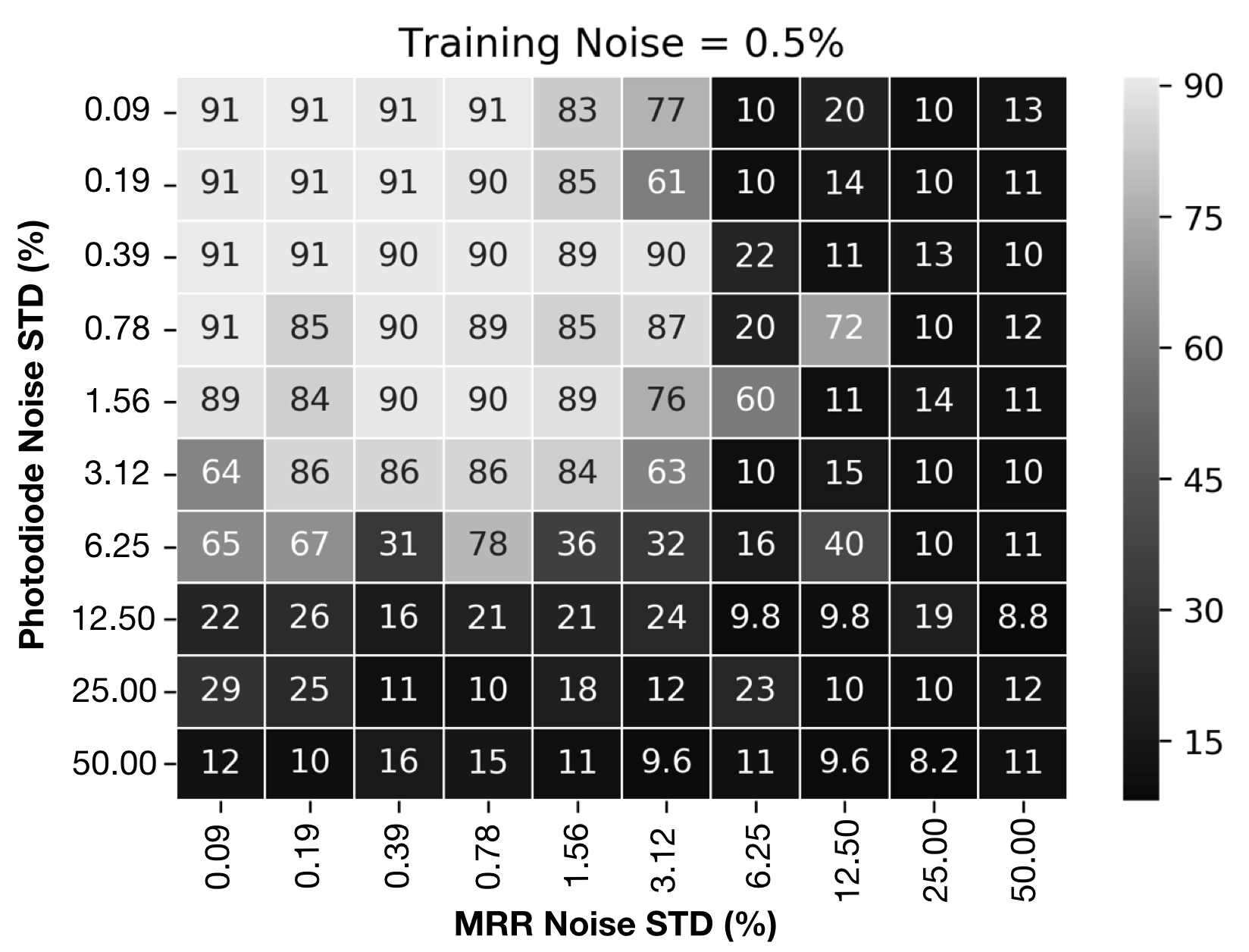}
         
         \label{fig:five over x}
     \end{subfigure}
        \caption{The evaluation of the effect of physical photodiode and MRR noise on inference accuracy. This effect can be partially compensated through introduction of an artificial noise source during the training stage. At the absence of training noise source (top) inference accuracy is quickly deteriorated as we sweep the photodiode and MRR noise. By introducing an equivalent of 0.1\% guassian noise, the network becomes more robust to inference noise. Further increase in training noise level (bottom) hinders the network from proper training.}
        \label{fig:three graphs}
\end{figure}

We initially trained our neural network offline on a conventional digital computer. Later during the inference stage we loaded the trained weights into our in-house simulator, which is equipped with noise sources modeling. Our hypothesis was that inference on a noisy neural network would result in some loss in accuracy. This is mostly due to the fact that, the network used during the training is noise-less, with 32-bit floating point resolution, while during inference the weights all in a sudden face a noisy network. In other words, the network performing inference experiences unseen noise behavior that results in accuracy loss. We tested our hypothesis by sweeping a range of inference noise levels and observing its effect on accuracy. For that reason, we identified two major noise sources, namely the neuron output noise and the weight noise. The neuron output noise represents the noise introduced at the output of each neuron by the photodiode and the nonlinear activation function. The first plot in Figure \ref{fig:three graphs}\subref{fig:no_train_noise} shows how accuracy is impacted by noise during inference for the case that the network was trained free of any noise source.\\

Our next hypothesis was that, if we allow for certain amount of noise during the training, the model would become more robust to noise during the inference stage. To that end, we trained the network with output noise source on. We only added the output noise, and left the weight noise off, because weights are required to be calculated with maximum precision during training. In fact we observed that even a minute amount of noise added to the weights during the training could destroy the accuracy of the network to its baseline level of about 10\%. We swept the addition of training noise at logarithmic steps from $0.01\%$ to $1\%$. Figure \ref{fig:three graphs}\subref{fig:noise_001} depicts the effect of adding an output noise equivalent to $0.1\%$ and $0.5\%$ of the maximum signal swing at the output of neurons. In our experiments we observed that the addition of about $0.1\%$ noise during the training may result in slight $2\%$ accuracy loss for low level of noise during inference. However, the model becomes more robust to higher levels of inference noise. This shows that modeling noise by addition of noise during training can fine-tune the network for a physical noisy realization as shown in Figure \ref{fig:three graphs} (middle). Lastly, we noticed adding further amount of training noise beyond the initial $0.1\%$ resulted very significant inference accuracy losses shown in Figure \ref{fig:three graphs} (bottom). 
\section{Conclusion} In this paper we presented a photonic CNN accelerator based on Winograd filering convolution algorithm. Winograd reduces the total number of multiplication, thus hardware, to perform convolution operation. We evaluated the speed of our accelerator by developing an analyical framework. Our results show that a photonic accelerator can  compete with state-of-the-art Winograd based FPGA and GPU implementations. Such photonic accelerator has the potential of improving the energy efficiency by up to three orders of magnitude. However, the overall speed is bound by the limitations of IO and conversions in DAC and ADC. To evaluate power performance we augmented the native hardware-agnostic Google Tensorflow tool with power models of our hardware components. Similar to speed performance, electronic IO and convertors are the major consumers of power in our proposed design. However, the photonic core, without the electronic interface, can operate while consuming up to two orders of magnitude less power. In addition, we modeled noise into our Tensorflow-based simulator, to investigate the effect of hardware noise sources such as photodiode noise and MRR noise on the functionality (accuracy) of our CNN. We found training the CNN with a small noise component, $0.1\%$ of the signal swing in our experiment, can result in the CNN become more robust to inference-time noise introduced by noisy photodiodes and MRRs.


%





\ifCLASSOPTIONcaptionsoff
  \newpage
\fi



%
\bibliographystyle{IEEEtran}
\bibliography{./refs.bib}

\begin{thebibliography}{10}
\providecommand{\url}[1]{#1}
\csname url@samestyle\endcsname
\providecommand{\newblock}{\relax}
\providecommand{\bibinfo}[2]{#2}
\providecommand{\BIBentrySTDinterwordspacing}{\spaceskip=0pt\relax}
\providecommand{\BIBentryALTinterwordstretchfactor}{4}
\providecommand{\BIBentryALTinterwordspacing}{\spaceskip=\fontdimen2\font plus
\BIBentryALTinterwordstretchfactor\fontdimen3\font minus
  \fontdimen4\font\relax}
\providecommand{\BIBforeignlanguage}[2]{{%
\expandafter\ifx\csname l@#1\endcsname\relax
\typeout{** WARNING: IEEEtran.bst: No hyphenation pattern has been}%
\typeout{** loaded for the language `#1'. Using the pattern for}%
\typeout{** the default language instead.}%
\else
\language=\csname l@#1\endcsname
\fi
#2}}
\providecommand{\BIBdecl}{\relax}
\BIBdecl

\bibitem{prucnal_neuromorphic_2017}
P.~R. Prucnal and B.~J. Shastri, \emph{\BIBforeignlanguage{en}{Neuromorphic
  {Photonics}}}.\hskip 1em plus 0.5em minus 0.4em\relax CRC Press, May 2017,
  google-Books-ID: VbvODgAAQBAJ.

\bibitem{chakraborty2018toward}
I.~Chakraborty, G.~Saha, A.~Sengupta, and K.~Roy, ``Toward fast neural
  computing using all-photonic phase change spiking neurons,'' \emph{Scientific
  reports}, vol.~8, no.~1, p. 12980, 2018.

\bibitem{feldmann2019all}
J.~Feldmann, N.~Youngblood, C.~Wright, H.~Bhaskaran, and W.~Pernice,
  ``All-optical spiking neurosynaptic networks with self-learning
  capabilities,'' \emph{Nature}, vol. 569, no. 7755, p. 208, 2019.

\bibitem{george_neuromorphic_2019}
\BIBentryALTinterwordspacing
J.~K. George, A.~Mehrabian, R.~Amin, J.~Meng, T.~F.~d. Lima, A.~N. Tait, B.~J.
  Shastri, T.~El-Ghazawi, P.~R. Prucnal, and V.~J. Sorger,
  ``\BIBforeignlanguage{EN}{Neuromorphic photonics with electro-absorption
  modulators},'' \emph{\BIBforeignlanguage{EN}{Optics Express}}, vol.~27,
  no.~4, pp. 5181--5191, Feb. 2019. [Online]. Available:
  \url{https://www.osapublishing.org/oe/abstract.cfm?uri=oe-27-4-5181}
\BIBentrySTDinterwordspacing

\bibitem{wang_integrated_2018}
\BIBentryALTinterwordspacing
C.~Wang, M.~Zhang, X.~Chen, M.~Bertrand, A.~Shams-Ansari, S.~Chandrasekhar,
  P.~Winzer, and M.~LonÄar, ``\BIBforeignlanguage{En}{Integrated lithium
  niobate electro-optic modulators operating at {CMOS}-compatible voltages},''
  \emph{\BIBforeignlanguage{En}{Nature}}, vol. 562, no. 7725, p. 101, Oct.
  2018. [Online]. Available:
  \url{https://www.nature.com/articles/s41586-018-0551-y}
\BIBentrySTDinterwordspacing

\bibitem{liu_high-speed_2004}
\BIBentryALTinterwordspacing
A.~Liu, R.~Jones, L.~Liao, D.~Samara-Rubio, D.~Rubin, O.~Cohen, R.~Nicolaescu,
  and M.~Paniccia, ``\BIBforeignlanguage{En}{A high-speed silicon optical
  modulator based on a metalâoxideâsemiconductor capacitor},''
  \emph{\BIBforeignlanguage{En}{Nature}}, vol. 427, no. 6975, p. 615, Feb.
  2004. [Online]. Available: \url{https://www.nature.com/articles/nature02310}
\BIBentrySTDinterwordspacing

\bibitem{amin_0.52_2018}
\BIBentryALTinterwordspacing
R.~Amin, R.~Maiti, C.~Carfano, Z.~Ma, M.~H. Tahersima, Y.~Lilach, D.~Ratnayake,
  H.~Dalir, and V.~J. Sorger, ``0.52 {V} mm {ITO}-based {Mach}-{Zehnder}
  modulator in silicon photonics,'' \emph{APL Photonics}, vol.~3, no.~12, p.
  126104, Dec. 2018. [Online]. Available:
  \url{https://aip.scitation.org/doi/10.1063/1.5052635}
\BIBentrySTDinterwordspacing

\bibitem{bagherian2018chip}
H.~Bagherian, S.~Skirlo, Y.~Shen, H.~Meng, V.~Ceperic, and M.~Soljacic,
  ``On-chip optical convolutional neural networks,'' \emph{arXiv preprint
  arXiv:1808.03303}, 2018.

\bibitem{mehrabian2018pcnna}
A.~Mehrabian, Y.~Al-Kabani, V.~J. Sorger, and T.~El-Ghazawi, ``Pcnna: A
  photonic convolutional neural network accelerator,'' in \emph{2018 31st IEEE
  International System-on-Chip Conference (SOCC)}.\hskip 1em plus 0.5em minus
  0.4em\relax IEEE, 2018, pp. 169--173.

\bibitem{liu2019holylight}
W.~Liu, W.~Liu, Y.~Ye, Q.~Lou, Y.~Xie, and L.~Jiang, ``Holylight: A
  nanophotonic accelerator for deep learning in data centers,'' in \emph{2019
  Design, Automation \& Test in Europe Conference \& Exhibition (DATE)}.\hskip
  1em plus 0.5em minus 0.4em\relax IEEE, 2019, pp. 1483--1488.

\bibitem{noauthor_optalysys_nodate}
\BIBentryALTinterwordspacing
``\BIBforeignlanguage{en-GB}{Optalysys}.'' [Online]. Available:
  \url{https://www.optalysys.com/}
\BIBentrySTDinterwordspacing

\bibitem{shen_deep_2017}
\BIBentryALTinterwordspacing
Y.~Shen, N.~C. Harris, S.~Skirlo, M.~Prabhu, T.~Baehr-Jones, M.~Hochberg,
  X.~Sun, S.~Zhao, H.~Larochelle, D.~Englund, and M.~SoljaÄiÄ,
  ``\BIBforeignlanguage{en}{Deep learning with coherent nanophotonic
  circuits},'' \emph{\BIBforeignlanguage{en}{Nature Photonics}}, vol.~11,
  no.~7, pp. 441--446, Jul. 2017. [Online]. Available:
  \url{http://www.nature.com/articles/nphoton.2017.93}
\BIBentrySTDinterwordspacing

\bibitem{hughes_training_2018}
\BIBentryALTinterwordspacing
T.~W. Hughes, M.~Minkov, Y.~Shi, and S.~Fan, ``\BIBforeignlanguage{EN}{Training
  of photonic neural networks through in situ backpropagation and gradient
  measurement},'' \emph{\BIBforeignlanguage{EN}{Optica}}, vol.~5, no.~7, pp.
  864--871, Jul. 2018. [Online]. Available:
  \url{https://www.osapublishing.org/optica/abstract.cfm?uri=optica-5-7-864}
\BIBentrySTDinterwordspacing

\bibitem{miscuglio_all-optical_2018}
\BIBentryALTinterwordspacing
M.~Miscuglio, A.~Mehrabian, Z.~Hu, S.~I. Azzam, J.~George, A.~V. Kildishev,
  M.~Pelton, and V.~J. Sorger, ``\BIBforeignlanguage{en}{All-optical nonlinear
  activation function for photonic neural networks [{Invited}]},''
  \emph{\BIBforeignlanguage{en}{Optical Materials Express}}, vol.~8, no.~12, p.
  3851, Dec. 2018. [Online]. Available:
  \url{https://www.osapublishing.org/abstract.cfm?URI=ome-8-12-3851}
\BIBentrySTDinterwordspacing

\bibitem{lavin2016fast}
A.~Lavin and S.~Gray, ``Fast algorithms for convolutional neural networks,'' in
  \emph{Proceedings of the IEEE Conference on Computer Vision and Pattern
  Recognition}, 2016, pp. 4013--4021.

\bibitem{lu2017evaluating}
L.~Lu, Y.~Liang, Q.~Xiao, and S.~Yan, ``Evaluating fast algorithms for
  convolutional neural networks on fpgas,'' in \emph{2017 IEEE 25th Annual
  International Symposium on Field-Programmable Custom Computing Machines
  (FCCM)}.\hskip 1em plus 0.5em minus 0.4em\relax IEEE, 2017, pp. 101--108.

\bibitem{shen2017deep}
Y.~Shen, N.~C. Harris, S.~Skirlo, M.~Prabhu, T.~Baehr-Jones, M.~Hochberg,
  X.~Sun, S.~Zhao, H.~Larochelle, D.~Englund \emph{et~al.}, ``Deep learning
  with coherent nanophotonic circuits,'' \emph{Nature Photonics}, vol.~11,
  no.~7, p. 441, 2017.

\bibitem{tait2014broadcast}
A.~N. Tait, M.~A. Nahmias, B.~J. Shastri, and P.~R. Prucnal, ``Broadcast and
  weight: an integrated network for scalable photonic spike processing,''
  \emph{Journal of Lightwave Technology}, vol.~32, no.~21, pp. 3427--3439,
  2014.

\bibitem{tait2016microring}
A.~N. Tait, A.~X. Wu, T.~F. de~Lima, E.~Zhou, B.~J. Shastri, M.~A. Nahmias, and
  P.~R. Prucnal, ``Microring weight banks,'' \emph{IEEE Journal of Selected
  Topics in Quantum Electronics}, vol.~22, no.~6, pp. 312--325, 2016.

\bibitem{hu2018single}
H.~Hu, F.~Da~Ros, M.~Pu, F.~Ye, K.~Ingerslev, E.~P. da~Silva, M.~Nooruzzaman,
  Y.~Amma, Y.~Sasaki, T.~Mizuno \emph{et~al.}, ``Single-source chip-based
  frequency comb enabling extreme parallel data transmission,'' \emph{Nature
  Photonics}, vol.~12, no.~8, p. 469, 2018.

\bibitem{xu2008silicon}
Q.~Xu, D.~Fattal, and R.~G. Beausoleil, ``Silicon microring resonators with
  1.5-$\mu$m radius,'' \emph{Optics express}, vol.~16, no.~6, pp. 4309--4315,
  2008.

\bibitem{chen2009integrated}
L.~Chen, K.~Preston, S.~Manipatruni, and M.~Lipson, ``Integrated ghz silicon
  photonic interconnect with micrometer-scale modulators and detectors,''
  \emph{Optics express}, vol.~17, no.~17, pp. 15\,248--15\,256, 2009.

\bibitem{stathopoulos2017multibit}
S.~Stathopoulos, A.~Khiat, M.~Trapatseli, S.~Cortese, A.~Serb, I.~Valov, and
  T.~Prodromakis, ``Multibit memory operation of metal-oxide bi-layer
  memristors,'' \emph{Scientific reports}, vol.~7, no.~1, p. 17532, 2017.

\bibitem{xu201723}
B.~Xu, Y.~Zhou, and Y.~Chiu, ``A 23-mw 24-gs/s 6-bit voltage-time hybrid
  time-interleaved adc in 28-nm cmos,'' \emph{IEEE Journal of Solid-State
  Circuits}, vol.~52, no.~4, pp. 1091--1100, 2017.

\bibitem{ziebell_ten_2011}
\BIBentryALTinterwordspacing
M.~Ziebell, D.~Marris-Morini, G.~Rasigade, P.~Crozat, J.-M. FÃ©dÃ©li,
  P.~Grosse, E.~Cassan, and L.~Vivien, ``\BIBforeignlanguage{EN}{Ten {Gbit}/s
  ring resonator silicon modulator based on interdigitated {PN} junctions},''
  \emph{\BIBforeignlanguage{EN}{Optics Express}}, vol.~19, no.~15, pp.
  14\,690--14\,695, Jul. 2011. [Online]. Available:
  \url{https://www.osapublishing.org/oe/abstract.cfm?uri=oe-19-15-14690}
\BIBentrySTDinterwordspacing

\bibitem{gardes_high-speed_2009}
\BIBentryALTinterwordspacing
F.~Y. Gardes, A.~Brimont, P.~Sanchis, G.~Rasigade, D.~Marris-Morini,
  L.~O'Faolain, F.~Dong, J.~M. Fedeli, P.~Dumon, L.~Vivien, T.~F. Krauss, G.~T.
  Reed, and J.~MartÃ­, ``\BIBforeignlanguage{EN}{High-speed modulation of a
  compact silicon ring resonator based on a reverse-biased pn diode},''
  \emph{\BIBforeignlanguage{EN}{Optics Express}}, vol.~17, no.~24, pp.
  21\,986--21\,991, Nov. 2009. [Online]. Available:
  \url{https://www.osapublishing.org/oe/abstract.cfm?uri=oe-17-24-21986}
\BIBentrySTDinterwordspacing

\bibitem{noauthor_osa_nodate}
\BIBentryALTinterwordspacing
``{OSA} {\textbar} {Ten} {Gbit}/s ring resonator silicon modulator based on
  interdigitated {PN} junctions.'' [Online]. Available:
  \url{https://www.osapublishing.org/oe/abstract.cfm?uri=oe-19-15-14690}
\BIBentrySTDinterwordspacing

\bibitem{baba_50-gb/s_2013}
\BIBentryALTinterwordspacing
T.~Baba, S.~Akiyama, M.~Imai, N.~Hirayama, H.~Takahashi, Y.~Noguchi,
  T.~Horikawa, and T.~Usuki, ``\BIBforeignlanguage{EN}{50-{Gb}/s
  ring-resonator-based silicon modulator},''
  \emph{\BIBforeignlanguage{EN}{Optics Express}}, vol.~21, no.~10, pp.
  11\,869--11\,876, May 2013. [Online]. Available:
  \url{https://www.osapublishing.org/oe/abstract.cfm?uri=oe-21-10-11869}
\BIBentrySTDinterwordspacing

\bibitem{dong_low_2009}
\BIBentryALTinterwordspacing
P.~Dong, S.~Liao, D.~Feng, H.~Liang, D.~Zheng, R.~Shafiiha, C.-C. Kung,
  W.~Qian, G.~Li, X.~Zheng, A.~V. Krishnamoorthy, and M.~Asghari,
  ``\BIBforeignlanguage{en}{Low {V}\_pp, ultralow-energy, compact, high-speed
  silicon electro-optic modulator},'' \emph{\BIBforeignlanguage{en}{Optics
  Express}}, vol.~17, no.~25, p. 22484, Dec. 2009. [Online]. Available:
  \url{https://www.osapublishing.org/oe/abstract.cfm?uri=oe-17-25-22484}
\BIBentrySTDinterwordspacing

\bibitem{jayatilleka_crosstalk_2016}
\BIBentryALTinterwordspacing
H.~Jayatilleka, K.~Murray, M.~Caverley, N.~A.~F. Jaeger, L.~Chrostowski, and
  S.~Shekhar, ``Crosstalk in {SOI} {Microring} {Resonator}-{Based} {Filters},''
  \emph{Journal of Lightwave Technology}, vol.~34, no.~12, pp. 2886--2896, Jun.
  2016. [Online]. Available: \url{http://ieeexplore.ieee.org/document/7272050/}
\BIBentrySTDinterwordspacing

\bibitem{bahadori_crosstalk_2016}
\BIBentryALTinterwordspacing
M.~Bahadori, S.~Rumley, H.~Jayatilleka, K.~Murray, N.~A.~F. Jaeger,
  L.~Chrostowski, S.~Shekhar, and K.~Bergman, ``Crosstalk {Penalty} in
  {Microring}-{Based} {Silicon} {Photonic} {Interconnect} {Systems},''
  \emph{Journal of Lightwave Technology}, vol.~34, no.~17, pp. 4043--4052, Sep.
  2016. [Online]. Available: \url{http://ieeexplore.ieee.org/document/7506337/}
\BIBentrySTDinterwordspacing

\bibitem{noauthor_imec-epixfab_nodate}
\BIBentryALTinterwordspacing
``imec-{ePIXfab} {SiPhotonics} {Passives}.'' [Online]. Available:
  \url{http://www.europractice-ic.com/SiPhotonics\_technology\_IHP\_passives.php}
\BIBentrySTDinterwordspacing

\bibitem{vivien_zero-bias_2012}
\BIBentryALTinterwordspacing
L.~Vivien, A.~Polzer, D.~Marris-Morini, J.~Osmond, J.~M. Hartmann, P.~Crozat,
  E.~Cassan, C.~Kopp, H.~Zimmermann, and J.~M. FÃ©dÃ©li,
  ``\BIBforeignlanguage{EN}{Zero-bias 40gbit/s germanium waveguide
  photodetector on silicon},'' \emph{\BIBforeignlanguage{EN}{Optics Express}},
  vol.~20, no.~2, pp. 1096--1101, Jan. 2012. [Online]. Available:
  \url{https://www.osapublishing.org/oe/abstract.cfm?uri=oe-20-2-1096}
\BIBentrySTDinterwordspacing

\bibitem{salamin_100_2018}
\BIBentryALTinterwordspacing
Y.~Salamin, P.~Ma, B.~Baeuerle, A.~Emboras, Y.~Fedoryshyn, W.~Heni, B.~Cheng,
  A.~Josten, and J.~Leuthold, ``\BIBforeignlanguage{en}{100 {GHz} {Plasmonic}
  {Photodetector}},'' \emph{\BIBforeignlanguage{en}{ACS Photonics}}, vol.~5,
  no.~8, pp. 3291--3297, Aug. 2018. [Online]. Available:
  \url{http://pubs.acs.org/doi/10.1021/acsphotonics.8b00525}
\BIBentrySTDinterwordspacing

\bibitem{ma_100_2018}
\BIBentryALTinterwordspacing
P.~Ma, Y.~Salamin, B.~Baeuerle, A.~Emboras, Y.~Fedoryshyn, W.~Heni, B.~Cheng,
  A.~Josten, and J.~Leuthold, ``\BIBforeignlanguage{EN}{100 {GHz}
  {Photoconductive} {Plasmonic} {Germanium} {Detector}},'' in
  \emph{\BIBforeignlanguage{EN}{Conference on {Lasers} and {Electro}-{Optics}
  (2018), paper {SM}2I.3}}.\hskip 1em plus 0.5em minus 0.4em\relax Optical
  Society of America, May 2018, p. SM2I.3. [Online]. Available:
  \url{https://www.osapublishing.org/abstract.cfm?uri=CLEO\_SI-2018-SM2I.3}
\BIBentrySTDinterwordspacing

\bibitem{tait_neuromorphic_2017}
\BIBentryALTinterwordspacing
A.~N. Tait, T.~F. de~Lima, E.~Zhou, A.~X. Wu, M.~A. Nahmias, B.~J. Shastri, and
  P.~R. Prucnal, ``Neuromorphic photonic networks using silicon photonic weight
  banks,'' \emph{Scientific Reports}, vol.~7, no.~1, p. 7430, Aug. 2017.
  [Online]. Available: \url{https://doi.org/10.1038/s41598-017-07754-z}
\BIBentrySTDinterwordspacing

\bibitem{bogaerts2012silicon}
W.~Bogaerts, P.~De~Heyn, T.~Van~Vaerenbergh, K.~De~Vos, S.~Kumar~Selvaraja,
  T.~Claes, P.~Dumon, P.~Bienstman, D.~Van~Thourhout, and R.~Baets, ``Silicon
  microring resonators,'' \emph{Laser \& Photonics Reviews}, vol.~6, no.~1, pp.
  47--73, 2012.

\bibitem{xue2016thermal}
X.~Xue, Y.~Xuan, C.~Wang, P.-H. Wang, Y.~Liu, B.~Niu, D.~E. Leaird, M.~Qi, and
  A.~M. Weiner, ``Thermal tuning of kerr frequency combs in silicon nitride
  microring resonators,'' \emph{Optics express}, vol.~24, no.~1, pp. 687--698,
  2016.

\bibitem{yoshida2007high}
C.~Yoshida, K.~Tsunoda, H.~Noshiro, and Y.~Sugiyama, ``High speed resistive
  switching in pt/ ti o 2/ ti n film for nonvolatile memory application,''
  \emph{Applied Physics Letters}, vol.~91, no.~22, p. 223510, 2007.

\bibitem{borghetti2010memristive}
J.~Borghetti, G.~S. Snider, P.~J. Kuekes, J.~J. Yang, D.~R. Stewart, and R.~S.
  Williams, ``‘memristive’switches enable ‘stateful’logic operations
  via material implication,'' \emph{Nature}, vol. 464, no. 7290, p. 873, 2010.

\bibitem{baek2004highly}
I.~Baek, M.~Lee, S.~Seo, M.~Lee, D.~Seo, D.-S. Suh, J.~Park, S.~Park, H.~Kim,
  I.~Yoo \emph{et~al.}, ``Highly scalable nonvolatile resistive memory using
  simple binary oxide driven by asymmetric unipolar voltage pulses,'' in
  \emph{IEDM Technical Digest. IEEE International Electron Devices Meeting,
  2004.}\hskip 1em plus 0.5em minus 0.4em\relax IEEE, 2004, pp. 587--590.

\bibitem{merced2016repeatable}
E.~J. Merced-Grafals, N.~D{\'a}vila, N.~Ge, R.~S. Williams, and J.~P. Strachan,
  ``Repeatable, accurate, and high speed multi-level programming of memristor
  1t1r arrays for power efficient analog computing applications,''
  \emph{Nanotechnology}, vol.~27, no.~36, p. 365202, 2016.

\bibitem{prakash2015resistance}
A.~Prakash, D.~Deleruyelle, J.~Song, M.~Bocquet, and H.~Hwang, ``Resistance
  controllability and variability improvement in a taox-based resistive memory
  for multilevel storage application,'' \emph{Applied Physics Letters}, vol.
  106, no.~23, p. 233104, 2015.

\bibitem{lee2012multi}
S.~R. Lee, Y.-B. Kim, M.~Chang, K.~M. Kim, C.~B. Lee, J.~H. Hur, G.-S. Park,
  D.~Lee, M.-J. Lee, C.~J. Kim \emph{et~al.}, ``Multi-level switching of
  triple-layered taox rram with excellent reliability for storage class
  memory,'' in \emph{2012 Symposium on VLSI Technology (VLSIT)}.\hskip 1em plus
  0.5em minus 0.4em\relax IEEE, 2012, pp. 71--72.

\bibitem{simonyan2014very}
K.~Simonyan and A.~Zisserman, ``Very deep convolutional networks for
  large-scale image recognition,'' \emph{arXiv preprint arXiv:1409.1556}, 2014.

\bibitem{he2016deep}
K.~He, X.~Zhang, S.~Ren, and J.~Sun, ``Deep residual learning for image
  recognition,'' in \emph{Proceedings of the IEEE conference on computer vision
  and pattern recognition}, 2016, pp. 770--778.

\bibitem{szegedy2016rethinking}
C.~Szegedy, V.~Vanhoucke, S.~Ioffe, J.~Shlens, and Z.~Wojna, ``Rethinking the
  inception architecture for computer vision,'' in \emph{Proceedings of the
  IEEE conference on computer vision and pattern recognition}, 2016, pp.
  2818--2826.

\bibitem{mathieu2013fast}
M.~Mathieu, M.~Henaff, and Y.~LeCun, ``Fast training of convolutional networks
  through ffts,'' \emph{arXiv preprint arXiv:1312.5851}, 2013.

\bibitem{vasilache2014fast}
N.~Vasilache, J.~Johnson, M.~Mathieu, S.~Chintala, S.~Piantino, and Y.~LeCun,
  ``Fast convolutional nets with fbfft: A gpu performance evaluation,''
  \emph{arXiv preprint arXiv:1412.7580}, 2014.

\bibitem{chetlur2014cudnn}
S.~Chetlur, C.~Woolley, P.~Vandermersch, J.~Cohen, J.~Tran, B.~Catanzaro, and
  E.~Shelhamer, ``cudnn: Efficient primitives for deep learning,'' \emph{arXiv
  preprint arXiv:1410.0759}, 2014.

\bibitem{zhang2015optimizing}
C.~Zhang, P.~Li, G.~Sun, Y.~Guan, B.~Xiao, and J.~Cong, ``Optimizing fpga-based
  accelerator design for deep convolutional neural networks,'' in
  \emph{Proceedings of the 2015 ACM/SIGDA International Symposium on
  Field-Programmable Gate Arrays}.\hskip 1em plus 0.5em minus 0.4em\relax ACM,
  2015, pp. 161--170.

\bibitem{qiu2016going}
J.~Qiu, J.~Wang, S.~Yao, K.~Guo, B.~Li, E.~Zhou, J.~Yu, T.~Tang, N.~Xu, S.~Song
  \emph{et~al.}, ``Going deeper with embedded fpga platform for convolutional
  neural network,'' in \emph{Proceedings of the 2016 ACM/SIGDA International
  Symposium on Field-Programmable Gate Arrays}.\hskip 1em plus 0.5em minus
  0.4em\relax ACM, 2016, pp. 26--35.

\bibitem{suda2016throughput}
N.~Suda, V.~Chandra, G.~Dasika, A.~Mohanty, Y.~Ma, S.~Vrudhula, J.-s. Seo, and
  Y.~Cao, ``Throughput-optimized opencl-based fpga accelerator for large-scale
  convolutional neural networks,'' in \emph{Proceedings of the 2016 ACM/SIGDA
  International Symposium on Field-Programmable Gate Arrays}.\hskip 1em plus
  0.5em minus 0.4em\relax ACM, 2016, pp. 16--25.

\bibitem{zhang2018caffeine}
C.~Zhang, G.~Sun, Z.~Fang, P.~Zhou, P.~Pan, and J.~Cong, ``Caffeine: Towards
  uniformed representation and acceleration for deep convolutional neural
  networks,'' \emph{IEEE Transactions on Computer-Aided Design of Integrated
  Circuits and Systems}, 2018.

\bibitem{nazemi20153}
A.~Nazemi, K.~Hu, B.~Catli, D.~Cui, U.~Singh, T.~He, Z.~Huang, B.~Zhang,
  A.~Momtaz, and J.~Cao, ``3.4 a 36gb/s pam4 transmitter using an 8b 18gs/s dac
  in 28nm cmos,'' in \emph{2015 IEEE International Solid-State Circuits
  Conference-(ISSCC) Digest of Technical Papers}.\hskip 1em plus 0.5em minus
  0.4em\relax IEEE, 2015, pp. 1--3.

\bibitem{lee20151}
D.~U. Lee, K.~W. Kim, K.~W. Kim, K.~S. Lee, S.~J. Byeon, J.~H. Kim, J.~H. Cho,
  J.~Lee, and J.~H. Chun, ``A 1.2 v 8 gb 8-channel 128 gb/s high-bandwidth
  memory (hbm) stacked dram with effective i/o test circuits,'' \emph{IEEE
  Journal of Solid-State Circuits}, vol.~50, no.~1, pp. 191--203, 2015.

\bibitem{aimspdk2019}
\BIBentryALTinterwordspacing
``Silicon photonics process design kit (apsuny pdkv3.0),''
  http://http://www.aimphotonics.com/pdk. [Online]. Available:
  \url{http://www.aimphotonics.com/pdk}
\BIBentrySTDinterwordspacing

\end{thebibliography}
\addtolength{\textheight}{-15.cm} 

%

\vskip 0pt plus -1fil
\begin{IEEEbiography}[{\includegraphics[width=1.1in,height=1.25in,clip,keepaspectratio]{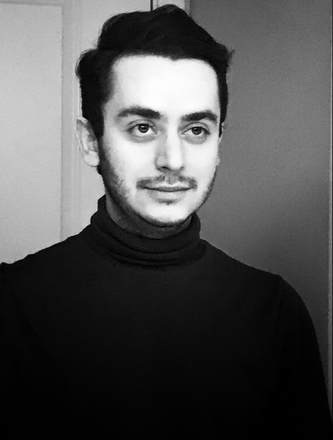}}]{Armin Mehrabian}
is a PhD candidate in Electrical Engineering at the George Washington University. His research interests include High Performance Computing (HPC), Neuromorphic Computing, Artificial Intelligence (AI) from both software and hardware point of view. He received his BS. degree in Electrical Engineering at Shahid Beheshti University of Tehran, Iran focusing on Analog Electronics, and his MS. degree at the George Washington University (GWU), DC, USA in computer engineering focusing on VLSI and digital electronics design. His current research involves leveraging nanophotonics for HPC architecture designs.
\end{IEEEbiography}
\vskip 0pt plus -1fil
\begin{IEEEbiography}[{\includegraphics[width=1.1in,height=1.25in,clip,keepaspectratio]{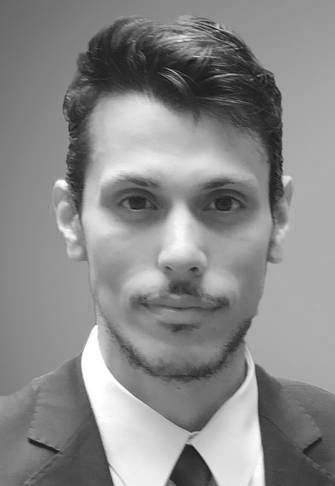}}]{Mario Miscuglio}
Mario Miscuglio is a post-doctoral researcher in the Electrical Engineering department at the George Washington University. He received his Masters’ in Electric and Computer engineering from Polytechnic of Turin, working as researcher at Harvard/MIT. He completed his PhD in Optoelectronics from University of Genova (IIT), working as research fellow at the Molecular Foundry in LBNL. His interests extend across science and engineering, including photonic neuromorphic computing, nano-optics and plasmonics. 
\end{IEEEbiography}
\vskip 0pt plus -1fil
\begin{IEEEbiography}[{\includegraphics[width=1.1in,height=1.25in,clip,keepaspectratio]{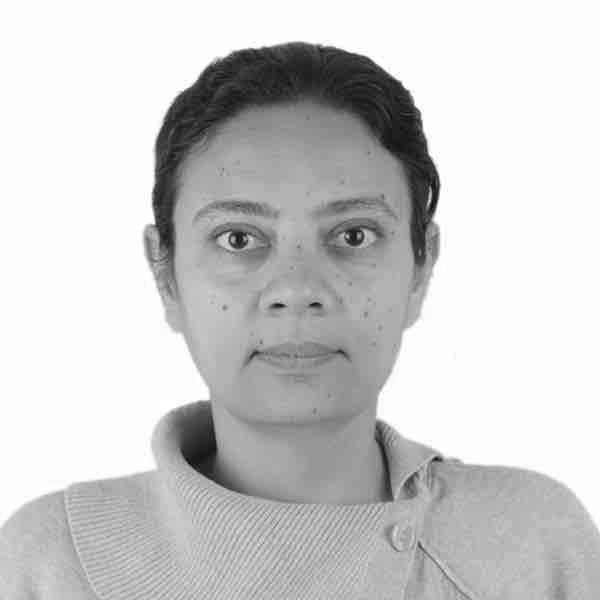}}]{Yousra Alkabani} received the BSc and MSc degrees in computer and systems engineering from Ain Shams University, Cairo, Egypt, in 2003 and 2006, respectively. She received the PhD degree in computer science from Rice University, Houston, TX, USA, in December 2010. She has been an assistant professor of computer and systems engineering at Ain Shams University since May 2011 and a visiting assistant professor of computer science and engineering at the American University in Cairo since 2013. Her research interests include hardware security, low power design, and embedded systems. She is a member of the IEEE.
\end{IEEEbiography}
\vskip 0pt plus -1fil

\begin{IEEEbiography}[{\includegraphics[width=1.1in,height=1.25in,clip,keepaspectratio]{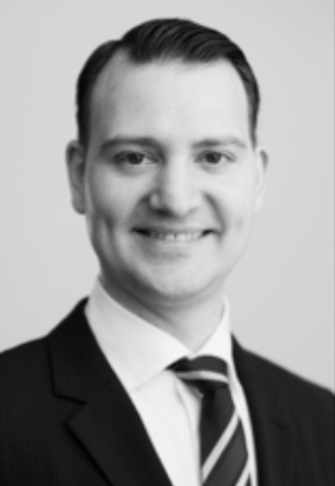}}]{Volker J. Sorger} is an Associate Professor in the Department of Electrical and Computer Engineering and the leader of the Orthogonal Physics Enabled Nanophotonics (OPEN) lab at the George Washington University. He received his PhD from the University of California Berkeley. His research areas include opto-electronic devices, plasmonics and nanophotonics and photonic analog information processing and neuromorphic computing. Amongst his breakthroughs are the first demonstration of a semiconductor plasmon laser, attojoule-efficient modulators, and PMAC/s-fast photonic neural networks and near real-time analog signal processors. Dr. Sorger has received multiple awards among are the Presidential Early Career Award for Scientists and Engineers (PECASE), the AFOSR Young Investigator Award (YIP), the Hegarty Innovation Prize, and the National Academy of Sciences award of the year. Dr. Sorger is the editor-in-chief of Nanophotonics, the OSA Division Chair for ‘Photonics and Opto-electronics’ and serves at the board-of-meetings at OSA \& SPIE, and the scholarship committee. He is a senior member of IEEE, OSA \& SPIE.
\end{IEEEbiography}
\vskip 0pt plus -20fil
\begin{IEEEbiography}[{\includegraphics[width=1in,height=1.25in,clip,keepaspectratio]{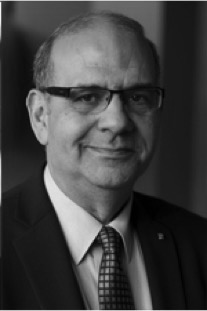}}]{Tarek El-Ghazawi}
is a Professor in the Department of Electrical and Computer Engineering at The George Washington University, where he leads the university-wide Strategic Program in High- Performance Computing. He is the founding director of The GW Institute for Massively Parallel Applications and Computing Technologies (IMPACT). His research interests include high-performance computing, parallel computer architectures, high-performance I/O, reconfigurable computing, experimental performance evaluations, computer vision, and remote sensing. He has published over 200 refereed research papers and book chapters in these areas and his research has been supported by DoD/DARPA, NASA, NSF, and also industry, including IBM and SGI. He is the first author of the book UPC: Distributed Shared Memory Programming, which has the first formal specification of the UPC language used in high-performance computing. Dr. El-Ghazawi is a member of the ACM and the Phi Kappa Phi National Honor Society; he was also a U.S. Fulbright Scholar, a recipient of the Alexander Schwarzkopf Prize for Technological Innovations and a recipient of the Alexander von Humboldt research award from the Humboldt Foundation in Germany. He is a fellow of the IEEE.
\end{IEEEbiography}




\end{document}